\documentclass[pre,aps,superscriptaddress,showpacs,preprint]{revtex4-1}

\usepackage{amssymb,amsmath}
\usepackage{graphicx}
\usepackage{nicefrac}
\usepackage{color}
\usepackage{gensymb}

\newcommand{\order}{{\cal O}}

\begin{document}

\title
{Microbranching in mode-I fracture using large scale simulations of amorphous and perturbed lattice models}
\author
{Shay I. Heizler}
\email{highzlers@walla.co.il}
\affiliation{Department of Physics, Bar-Ilan University, Ramat-Gan, IL52900 ISRAEL}
\affiliation{Department of Physics, Nuclear Research Center-Negev, P.O. Box 9001, Beer Sheva 84190, ISRAEL}
\author
{David A. Kessler}
\affiliation{Department of Physics, Bar-Ilan University, Ramat-Gan, IL52900 ISRAEL}
 

\begin{abstract}

We study the high-velocity regime mode-I fracture instability when small microbranches start to appear near the main crack, using large scale simulations.
Some of the features of those microbranches have been reproduced qualitatively in smaller
scale studies (using $\order(10^4)$ atoms) on both a model of an amorphous materials (via the continuous random network model) and using perturbed lattice models. In this study,
larger scale simulations ($\order(10^6)$ atoms) were performed using multi-threading
computing on a GPU device, in order to achieve more physically realistic results. First, we find that the microbranching pattern appears to be converging with the lattice width.
Second, the simulations reproduce the growth of the size of a microbranch as a function of the crack
velocity, as well as the increase of the amplitude of the derivative of the electrical resistance RMS with respect to the time as a function of the crack velocity.
In addition, the simulations yield the correct branching angle of the microbranches, and the power law governing the shape of the microbranches seems to be lower
than one, so that the side cracks turn over in the direction of propagation of the main crack as seen in experiment. 
 
\end{abstract}

\maketitle
 
\section{Introduction}

The study of the physics of brittle fracture has been very fruitful in the last two decades~\cite{review,review_mid,review_new}. New experiments have shown various new features
of dynamic fracture, focused in Mode-I (tensile) fracture using amorphous brittle materials~\cite{marder_jay2,fineberg_sharon2,fineberg_sharon5,livne}. In particular,
the experiments have shown a sharp transition between the regime of
steady-state cracks and the regime of unstable cracks~\cite{fineberg_sharon2,fineberg_sharon5}. In steady state, where the driving displacement, $\Delta$
(of order  $\Delta_G$, the Griffith criterion),
is sufficiently small, a single crack propagates in the midline of the sample, reaching  a steady-state velocity (which is 
of order the Rayleigh surface wave speed, $c_R$). Increasing $\Delta$ results in an increased steady-state velocity, yielding a  $v(\Delta)$ curve, until a specific critical point.
Increasing the diving displacement further, beyond this critical point, we enter the unstable regime, where small microbranches start to appear nearby the main crack.
The experiments have shown that above the critical point, the size of the average microbranch (which is log-normal distributed) increases rapidly with
the crack velocity, measured via the slope of the electrical resistance of a conductive layer that is pasted on the sample. The electrical
resistance slope exhibits oscillations whose amplitude increases rapidly as well. Increasing the driving displacement furthermore, the small microbranches
become large microbranches, creating a complex fracture pattern, and finally, creates macrobranches~\cite{review,fineberg_sharon5,ramulu_kobayashi}.

Some of the new experimental findings could not be explained via the classical theoretical approach for fracture mechanics, the linear elasticity fracture mechanics
(LEFM) theory~\cite{freund}. For example, several predictions of LEFM for the critical velocity, such as the studies of Yoffe~\cite{yoffe} or Eshelby~\cite{eshelby} predicted a single
universal critical velocity much higher than that seen in some of the experimental studies, such as in PMMA~\cite{review}. Also, experiments have found various material-dependent features
such as the terminal velocity which the crack manages to propagate as well as the critical velocity for
macro-branches~\cite{review}. As far as the micro-branching critical velocity, the question of universality is debatable~\cite{adda_bedia}, but anyhow,
the velocities are much smaller that the theoretical predictions.
For a review, see the Introduction in~\cite{shay4}. However, the basic reason for the failures of LEFM is that the basic equations
of LEFM yield a singularity of the stresses near the crack tip~\cite{freund},
and thus, the zone nearest  to the crack's tip (the process zone) cannot be modeled via LEFM. 

The failure of LEFM, caused by this singularity, gave rise to an interest in discrete lattice models and
simulations~\cite{slepyan,slepyan2,marderliu,kess_lev1,kess_lev3,pechenik,shay1,shay2,mdsim3,nature_old}. In this kind of model, the basic length-scale which is the
lattice scale, prevents the singularities that appear in the continuum approach. The lattice models were successful in reproducing the behavior of steady-state cracks~\cite{slepyan,slepyan2,kess_lev1},
including a material dependency of the $v(\Delta)$ curves, in that it depends on the specific parameters of the inter-atomic potential~\cite{shay1,shay2}.
Moreover, the lattice models predict a certain critical point beyond which the steady-state
solution becomes linearly unstable~\cite{kess_lev3,pechenik,shay1,shay2}. In the simulations, this is exactly the point which the crack stops propagating
along the midline of the sample and some additional bonds, not along the midline,
start to break~\cite{fineberg_mar,shay1,marderliu,kess_lev3}. However, especially in mode-I pure lattice simulations, the
post-instability behavior of the lattice models do not match the experiments, neither qualitatively or quantitatively~\cite{fineberg_mar,shay1}.

One recent approach to overcome these difficulties has been to turn to a more realistic model for an amorphous material,
the continuous random network (CRN) model~\cite{shay3}. The continuous random network was suggested
first by Zachariasen for describing amorphous material~\cite{zacharainsen}, and specific effective algorithms (using Monte-Carlo techniques)
for generating the CRN were offered in~\cite{www,www2,shay3}.
Recent accurate 2D experiments using transmission electron microscopy (TEM) in 2D silica on the structure of this amorphous material were reproduced to excellent
accuracy using the Zachariasen model~\cite{tem1,tem2}.
After generating an amorphous CRN sample, molecular
dynamics simulations were found to yield many of the important qualitative features of mode-I fracture experiments in amorphous brittle materials~\cite{shay3}.
The simulations showed the birth
of microbranches growing nearby the main crack, the increase of the size of the microbranches as a function of the driving displacement (or of course, the crack velocity), and the
growth of the amplitude of the derivative of the electrical resistance with respect to the time as a function of the crack velocity. 

Another direction recently examined was that of perturbed lattice models~\cite{shay4}, which 
exhibited  behavior similar to that of  the CRN model, including the main features mentioned above. However, these simulations (both the CRN and the perturbed lattice simulations)
suffered from a significant level of numerical noise, since each microbranch contained only
a few tens of broken bonds at most. Thus, the statistics that concerns the most interesting physics, that of the branches, was quite poor.

These intriguing results, performed on limited size systems ($\order(5\cdot10^4)$ atoms) provide strong motivation to conduct larger simulations
both to reduce the overall noise level and to get closer to at least a mesoscopic system where scaling behavior might set in.  The goal was to achieve
at least a two magnitude increase in size, i.e., simulations of the order of millions of atoms ($\order(5\cdot10^6)$ atoms), which necessitated using parallel (multi-threading) computing.
In this work we study mode-I fracture via large scale
simulations, with a particular focus  on the properties of the microbranches that could
not be studied in the previous, limited size, studies. In addition, recently experiments have been performed with gels, whose three orders of magnitude slower
sound speed (compared to the classic brittle materials like PMMA or glass)  enables direct visualization by means of moderately fast video cameras, resulting in  clear snapshots of the
crack tip~\cite{gels2,livne,review_new}. Using the larger scale simulations we can now compare the crack's tip shape, both on steady-state cracks in a strip geometry,
and especially, near the origin of the microbranching instability.

The models are presented in Sec. \ref{model}.
In Sec. \ref{validity} we perform some basic checks of our models, confirming that the transverse size of the microbranch zone ($\nicefrac{\delta y}{W}$) decreases with increasing
sample width $W$, for a given scaled driving displacement $\Delta/\Delta_G$ (in the experiments, using macroscopic sample sizes, the microbranching region width  is much smaller
than the sample width and the dynamics of the fracture 
is not effected by the sample edges). Next, in Sec. \ref{results} we present the quantitative results concerning the birth and the growth of the microbranches, and their physical features.
A short discussion is presented in Sec. \ref{discussion}. 

\section{Model and Main Methodology}
\label{model}

The simulations presented in this study are divided generally into two kinds. The first one uses the continuous random network (CRN)
model (to model an amorphous material), and the second employs a perturbed (honeycomb or hexagonal) lattice model.
Both models were described in depth in~\cite{shay3} and~\cite{shay4} respectively. We will review them here shortly. 

The CRN model  reproduces various structural features of real brittle
amorphous materials (though 2D, in this study) such as amorphous silicon or silica~\cite{tem1,tem2,laaziri}, and a 3D-extension of this model should reproduce the real behavior of fracture. The perturbed
honeycomb lattice is the ordered phase of the CRN, as discussed at length in~\cite{shay4} and shares similar features and results (though with less noise). In addition, in~\cite{tem1,tem2} there
is clear experimental 2D evidence that 2D ordered materials share the structural features of the perturbed honeycomb lattice. Also, the perturbed hexagonal lattice is a generalization of the hexagonal
lattice that was used in many studies to study dynamic fracture (for example~\cite{marderliu,review,nature_old}), and facilitates comparison with this body of work.    

\subsubsection{Generating the Continuous Random Network for modeling amorphous material}
We generated two-dimensional CRN's by a 2D-analogue~\cite{shay3} of the WWW algorithm~\cite{www,www2}.
The potential that was used in the construction of the CRN included both a 2-body central force and a 3-body bond-bending force:~\cite{potential,models}:
\begin{equation}
E_{\mathrm{tot}}=\sum_{i=1}^n\left[\sum_{j\in{\cal N}(i)}\frac{1}{4}k_r(\vert \vec{r_{ij}}\vert-a_{i,j}^2)+\sum_{j,k\in{\cal N}(i)}
\frac{1}{2}k_{\theta}(\cos\theta_{i,(j,k)}-\cos\theta_C)^2\right],
\label{potential}
\end{equation}
where $\vert \vec{r_{ij}}\vert$ is the radial distance between each pair of nearest-neighbor atoms and $a_{i,j}=a_0=4$ is a constant lattice scale (in contrast to the perturbed lattice model).
$k_r$ and $k_{\theta}$ are the radial and the azimuthal (3-body) spring constants, respectively.
$\cos\theta_{i,(j,k)}$ is the cosine of the angles between each set of 3 neighboring atoms, defined of course by:
\begin{equation}
\cos\theta_{i,j,k}=\frac{\vec{r}_{i,j}\cdot\vec{r}_{i,k}}{\vert\vec{r}_{i,j}\vert\vert\vec{r}_{i,k}\vert}
\label{cos_teta}
\end{equation}
where $i$ is the central atom and $(j,k)$ are its two neighbors. $\theta_C=\nicefrac{2\pi}{3}$ (characterizing a honeycomb lattice).
We start from a pure honeycomb lattice, randomize large
number of bonds and perform a Monte-Carlo procedure, wherein each cycle we switch two bonds,  calculating the optimal positions of the atoms in
the near zone of the switched bonds to determine the change of energy so as to decide whether to accept the switch.
Finally, we get a CRN that looks like Zachariasen's patterns~\cite{zacharainsen,shay3} and the TEM snapshots of the 2D amorphous Silica~\cite{tem1,tem2}.
For a in-depth discussion, see~\cite{shay3}. 

\subsubsection{Generating the perturbed lattice}
Here we start with a perfect honeycomb lattice and randomize the lattice scale of each ``bond", $a_{i,j}$:
\begin{equation}
a_{i,j}=(1+\epsilon_{i,j})a_0,\qquad i=1,2,\dots,n_{\mathrm{atoms}}, j\in {\cal N}(i)
\end{equation}
where $\epsilon_{i,j}\in[-b,b]$, and $b$ is constant for a given lattice, and in this work ranges between $0\leqslant b\leqslant 0.1$, $a_0=4$. ${\cal N}(i)$ refers
to the nearest-neighbors of site $i$. For a detailed discussion, see~\cite{shay4}. 

\subsubsection{The equations of motion}

In our model, between each two atoms there is a piece-wise linear radial force (2-body force law) of the form:
\begin{equation}
\vec{f}^r_{i,j}=k_rk'_{i,j}(\vert\vec{r}_{i,j}\vert-a_{i,j})\hat{r}_{j,i},
\label{force_Radial}
\end{equation}
where:
\begin{equation}
k'_{i,j}\equiv\theta_H\left(\varepsilon-\vert\vec{r}_{i,j}\vert\right).
\label{linear}
\end{equation}
The Heaviside step function $\theta_H$ guarantees that the force drops immediately to zero when the distance between two atoms $\vert\vec{r}_{i,j}\vert$ reaches
a certain value $\varepsilon>a_{i,j}$ (the breaking of a ``bond"). In this work we set $\varepsilon=a_0+1$. We describe here brittle materials with an extreme sharp transition
from linear response to failure, though, in reality the failure is always somewhat smoother.
Thus, alternatively to Eq. \ref{linear}, we can use a smoother force law,
which instead of a sharp failure at $\vert\vec{r}_{i,j}\vert=\varepsilon$, has a more realistic smooth transition wherein the force law drops to zero of the form~\cite{kess_lev3,shay1,shay2}:
\begin{equation}
k'_{i,j}\equiv\frac{1+\tanh [\alpha_{\mathrm{pot}} (\varepsilon -\vec{r}_{i,j})]}{1+\tanh (\alpha_{\mathrm{pot}})}
\label{nonlinear}
\end{equation}
where $\alpha_{\mathrm{pot}}$ is the smoothness parameter, such that when $\alpha_{\mathrm{pot}}\to\infty$  the force law reverts to the piecewise linear force law. The effect
of $\alpha_{\mathrm{pot}}$ on the fracture feature was investigated previously~\cite{kess_lev3,shay1,shay2}, and is reproduced in this paper. The results in this paper
 refer to  the piecewise linear model, unless mentioned otherwise.

In addition there is a 3-body force law that depends on the cosine of each of the angles, acts on the central atom (atom $i$) of each angle, and may be expressed as:
\begin{align}
\label{force_teta}
& \vec{f}^{\theta}_{i,(j,k)}=k_{\theta}(\cos\theta_{i,j,k}-\cos\theta_C)\frac{\partial\cos\theta_{i,j,k}}{\partial\vec{r}_i}\,
k'_{i,j}k'_{i,k}\hat{r}_i=\\
& k_{\theta}(\cos\theta_{i,j,k}-\cos\theta_C)\left[\frac{\vec{r}_{i,j}+\vec{r}_{i,k}}{\vert\vec{r}_{i,j}\vert\vert\vec{r}_{i,k}\vert}+
\frac{\vec{r}_{j,i}(\vec{r}_{i,j}\cdot\vec{r}_{i,k})}{\vert\vec{r}_{i,j}\vert^3\vert\vec{r}_{i,k}\vert}+
\frac{\vec{r}_{k,i}(\vec{r}_{i,j}\cdot\vec{r}_{i,k})}{\vert\vec{r}_{i,j}\vert\vert\vec{r}_{i,k}\vert^3} \right]k'_{i,j}k'_{i,k}, \nonumber
\end{align}
while the force that is applied on the other two atoms (atoms $j,k$) may expressed as:
\begin{align}
\label{force_teta2}
& \vec{f}^{\theta}_{j,(i,k)}=k_{\theta}(\cos\theta_{i,j,k}-\cos\theta_C)\frac{\partial\cos\theta_{i,j,k}}{\partial\vec{r}_j}
k'_{i,j}k'_{i,k}\hat{r}_j=\\
& k_{\theta}(\cos\theta_{i,j,k}-\cos\theta_C)\left[\frac{\vec{r}_{k,i}}{\vert\vec{r}_{i,j}\vert\vert\vec{r}_{i,k}\vert}+
\frac{\vec{r}_{i,j}(\vec{r}_{i,j}\cdot\vec{r}_{i,k})}{\vert\vec{r}_{i,j}\vert^3\vert\vec{r}_{i,k}\vert} \right]k'_{i,j}k'_{i,k} \nonumber
\end{align}
Of course, the forces satisfy the relation:
$\vec{f}^{\theta}_{i,(j,k)}=-(\vec{f}^{\theta}_{j,(i,k)}+\vec{f}^{\theta}_{k,(i,j)})$. The 3-body force law drops immediately to zero when using a piecewise linear force law
when the bond breaks (Eq. \ref{linear}), or may be taken to vanish smoothly, using Eq. \ref{nonlinear}.
In a honeycomb lattice there are three angles associated with each atom and in the hexagonal lattice there are six of them (we note that in the hexagonal lattice this choice is a little
bit arbitrary since there are in general additional optional angles for each atom). There is a certain preferred angle $\theta_C$ for which
the 3-body force law vanishes (in the honeycomb lattice we set $\theta_C=\nicefrac{2\pi}{3}$ and in the hexagonal lattice we set $\theta_C=\nicefrac{\pi}{3}$).

In addition, it is convenient to  add a small Kelvin-type viscoelastic force proportional to the relative velocity between the two atoms of
the bond $\vec{v}_{i,j}$~\cite{kess_lev1,kess_lev3,pechenik,shay1}:
\begin{equation}
\vec{g}^r_{i,j}=\eta(\vec{v}_{i,j}\cdot\hat{r}_{i,j})\,k'_{i,j}\hat{r}_{i,j},
\label{viscous}
\end{equation}
with $\eta$  the viscosity parameter. The viscous force also vanishes after the bond is broken, governed by $k'_{i,j}$. 
The imposition of a small amount of such a viscosity acts to stabilize the system and is especially useful in the relatively narrow systems simulated herein.

The set of equations of motion of each atom is then:
\begin{equation}
m_i\vec{\ddot{a}}_i=\sum_{j\in3p\;nn}\left(\vec{f}^r_{i,j}+\vec{g}^r_{i,j}\right)+\sum_{j,k\in3p\;nn}\vec{f}^{\theta}_{i,(j,k)}+\sum_{j\in6p\;nn}\vec{f}^{\theta}_{j,(i,k)},
\label{motion_equations}
\end{equation}
In this work the units are chosen so that the radial spring constant $k_r$ and the atoms mass $m_i$ is unity. We note that numerical measurement of the Young's modulus
of the hexagonal lattices for $k_{\theta}=0$ yields the well-known analytical  expression $2/\sqrt{3}k_r$ and is twice as
big ($E\approx2k_r$) with $k_{\theta}/k_r=10$ and the Poisson's ratio is $\nu=1/3$ as reported
in many previous works~\cite{youngs1,youngs2}. In the honeycomb lattice and the CRN (with $k_{\theta}/k_r=1$),
the Young's modulus is approximately $E\approx0.86k_r$ while the Poisson's ratio remains similar. However, the value of $k_r$ is not
crucial (in this study), since the results in this paper are in normalized units ($\Delta/\Delta_G$, $v/c_R$).

After relaxing the initial lattice, we strain the lattice under a mode-I tensile loading with a given constant strain corresponding a given driving displacement $\pm\Delta$ of the edges
(the top and bottom rows are held tight and do not allow transverse displacement) and seed the system with an initial crack (The left boundary condition is also held
in a pure ``cracked" state).
The crack then propagates via the same molecular dynamics Euler scheme using Eqs. \ref{force_Radial}-\ref{motion_equations}. We note that the calculation of $\Delta_G$ is set under
the equality of the energy in the uncracked uniform strain, to the energy needed for breaking the midline bonds in the sample (for example, see~\cite{kess_lev1} for the case of square lattice).
We parameterize our results in this paper using the normalized quantity $\Delta/\Delta_G$, but of course $\Delta/\Delta_G=K_I/K_{IC}$ (the stress intensity
factor normalized to the Griffith value)~\cite{review}.

\subsubsection{Parallelization by GPU computing}

As mentioned in the Introduction, the major innovation of this work, compared to our previous studies, is the use of large scale simulations. The previous studies of the amorphous (CRN)
 model~\cite{shay3} and the
perturbed lattice model~\cite{shay4} used approximately $50,000$ particles. in this study we wished to use approximately $5,000,000$ particles.
These kind of simulations cannot reasonably be performed by a singe CPU, and thus force us to use multi-thread computing.
We choose to use GPU computing, parallelizing the code via CUDA~\cite{padon,cooper}. This kind of programing forces the programmer to use the different levels of memory carefully~\cite{cooper},
which makes possible achieving an acceleration up to $\approx 100$ faster than a regular $C$ code using a single CPU. Beside the benefit of getting the results in a given system much faster,
the main benefit is the possibility to run large scale simulations, which was prohibitive before. This tool makes possible the simulation of millions
of atoms in  reasonable simulation times.

Our model consists of several modules,  each one of which needs to be re-written in CUDA. Both amorphous and lattice models use a molecular-dynamics module for
the fracture simulations that must be
parallelized. In addition, for the CRN, the Metropolis Monte-Carlo algorithm for generating the CRN needs to be parallelized. Furthermore,
the electrical resistance simulations, which we use to determine the crack velocity~\cite{shay3}, is solved by
a nonlinear Laplace solver that needs to be parallelized too. The electrical resistance is calculated via the method that was used in~\cite{shay3}, by solving the nonlinear Laplace on
a grid of resistors (each broken bond, in the main crack and in the microbranches is taken into account)~\cite{bonamy}. In Appendix \ref{appendix_a}
we discuss about the implementation of the different modules, the parameters that were used and the degree of acceleration achieved for the different modules with the GPU using CUDA.

The  various sized lattices we use contain:
\begin{itemize} 
\item {$162\cdot272\approx45,000$ ($N=80$ in the Slepyan model notation) atoms for the honeycomb lattice and $162\cdot408\approx65,000$ atoms for the hexagonal lattice, which we call $f=1$ (Factor=1).
This  size  is equal to  that used in our previous studies,~\cite{shay3} and~\cite{shay4}.}
\item {$486\cdot816\approx400,000$ ($N=240$ in the Slepyan model notation) atoms for the honeycomb lattice and $486\cdot1224\approx600,000$ atoms for the hexagonal lattice, which we call $f=3$ (Factor=3).}
\item {$1458\cdot2448\approx3,600,000$ ($N=720$ in the Slepyan model notation) atoms for the honeycomb lattice and $1296\cdot3264\approx4,200,000$ ($N=640$ in the Slepyan model notation)
atoms for the hexagonal lattice, which we call $f=8-9$ (Factor=9 in the honeycomb and CRN lattices and Factor=8 in the hexagonal lattice).}
\end{itemize}

In Appendix \ref{appendix_b} we present a brief discussion regarding the results for the CRN using the parallel GPU Monte-Carlo algorithm. The GPU algorithm reproduced the results
of the CPU code, in particular the agreement~\cite{shay3} of the radial distribution function with that experimentally determined~\cite{laaziri} for amorphous silicon.

\section{Overview of the Simulation Results}
\label{validity}

\begin{figure}
\centering{
(a)
\includegraphics*[width=7.5cm]{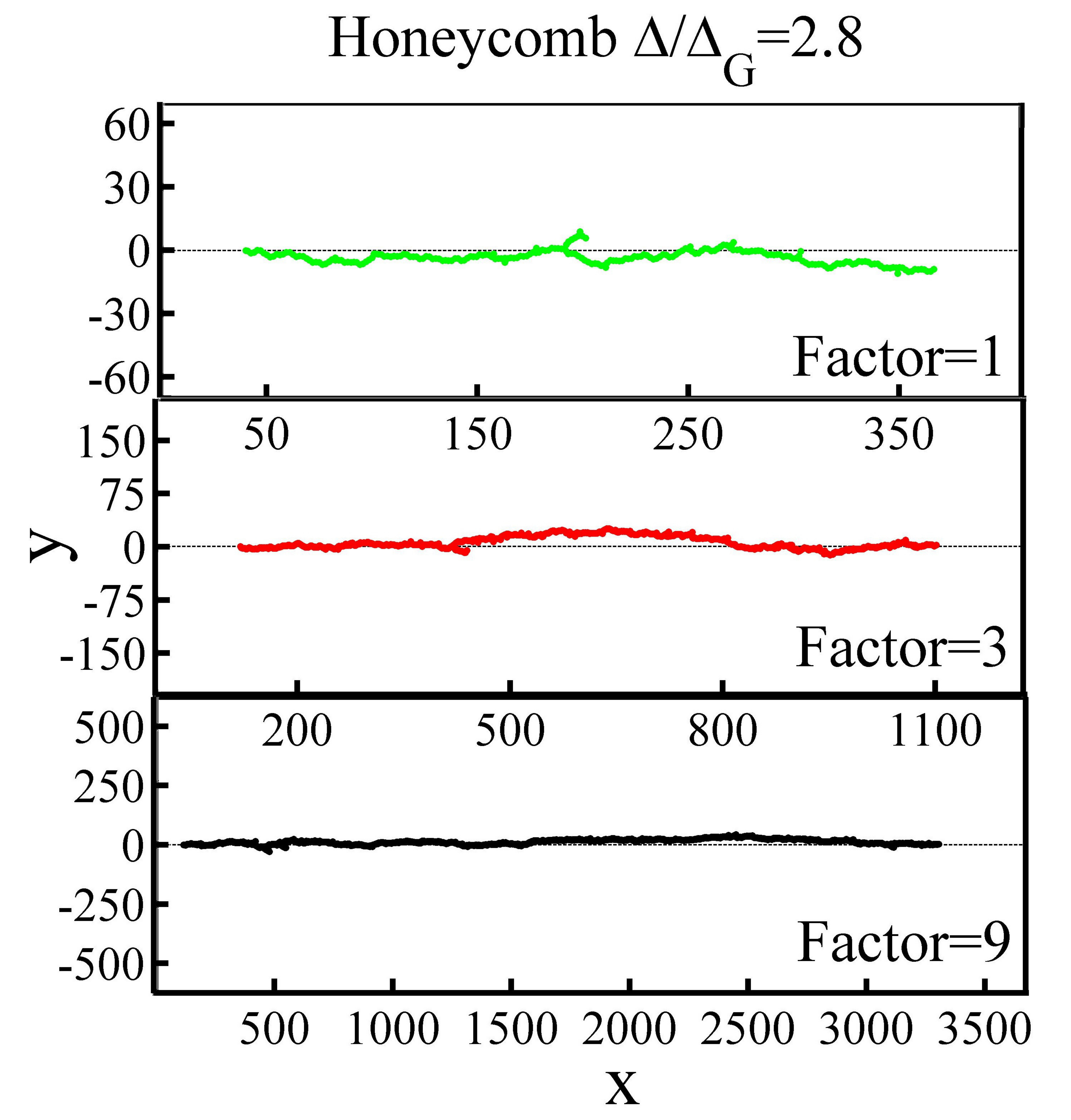}
(b)
\includegraphics*[width=7.5cm]{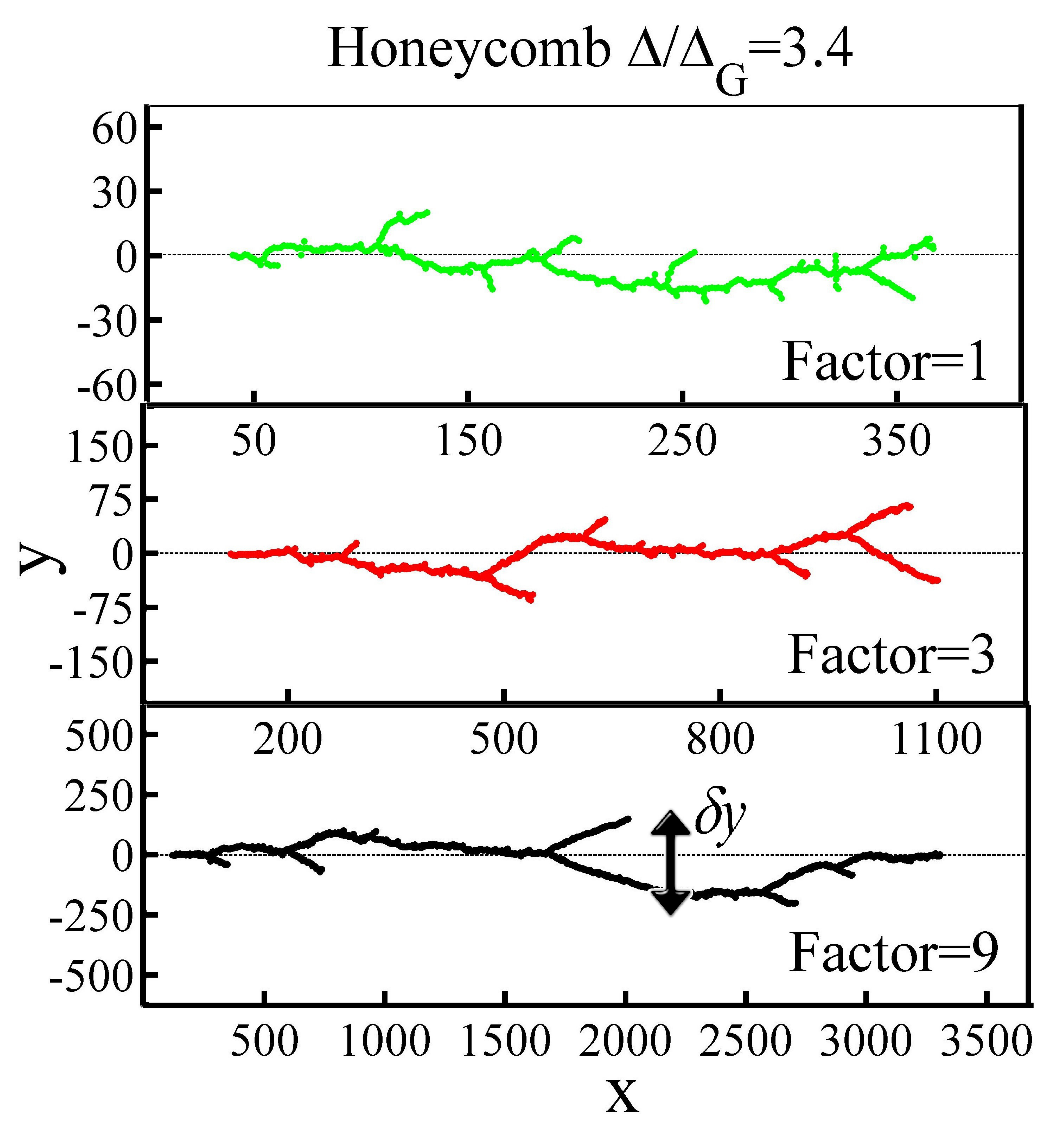}
(c)
\includegraphics*[width=7.5cm]{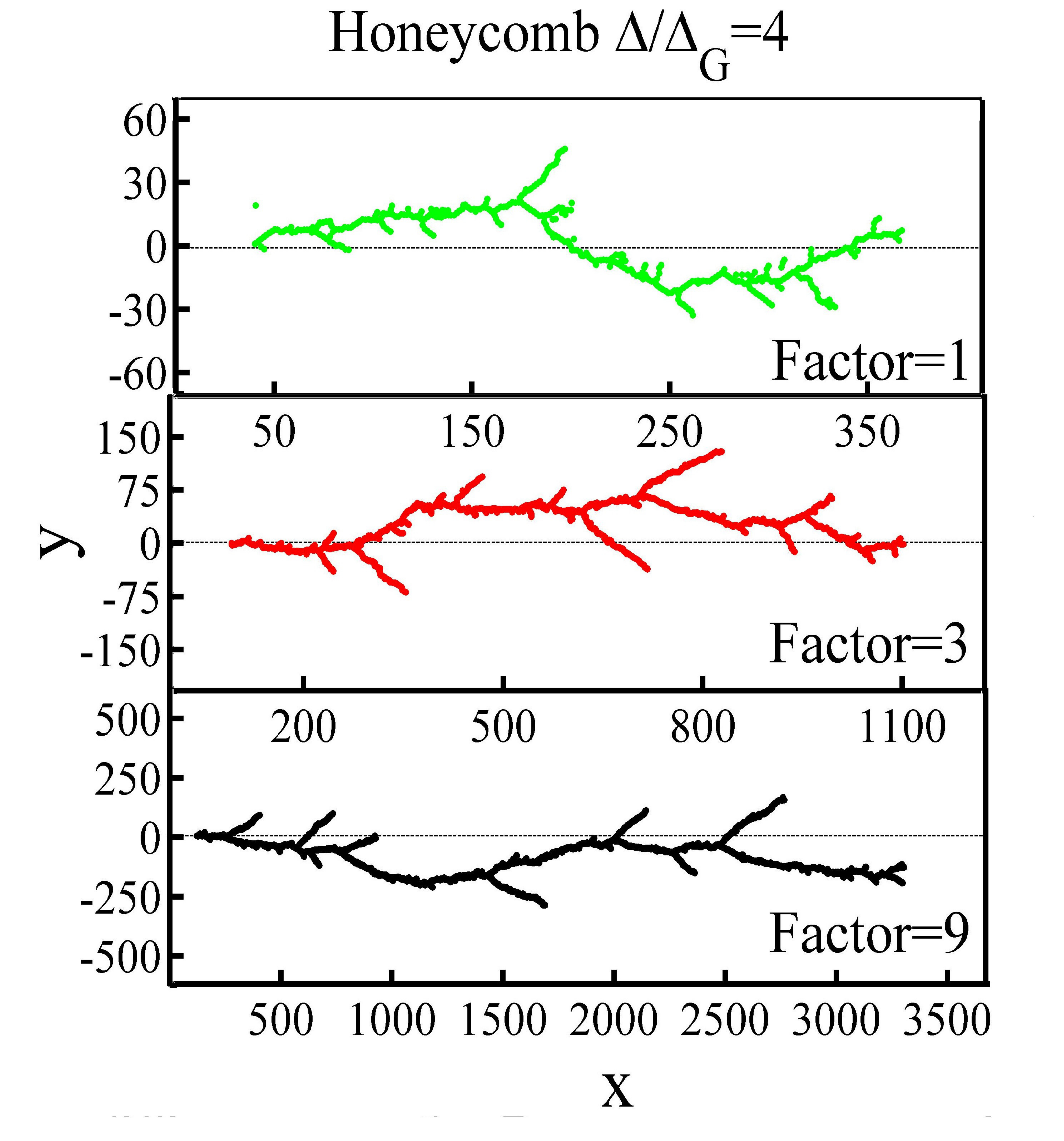}
}
\caption{(Color online) (a) The microbranching pattern in a perturbed honeycomb lattice for $\Delta/\Delta_G=2.8$ and $\eta=2$ for $f=1$ in the upper curve, 
for $f=3$ in the intermediate curve and for $f=9$ in the lower curve. $x$ and $y$ have the units of $a_0$, the lattice scale, and each figure depicts the whole sample
(except for the the initial seed crack that extends from starts from the left edge of each crack pattern to $x=y=0$). (b) The same for $\Delta/\Delta_G=3.4$. (c) The same for $\Delta/\Delta_G=4.0$.
In (b) we define $\delta y$, the width of the microbranching region, as the difference between the maximum and minimum $y$'s of broken bonds.}
\label{pattarns}
\end{figure}

\begin{figure}
\centering{
(a)
\includegraphics*[width=7.5cm]{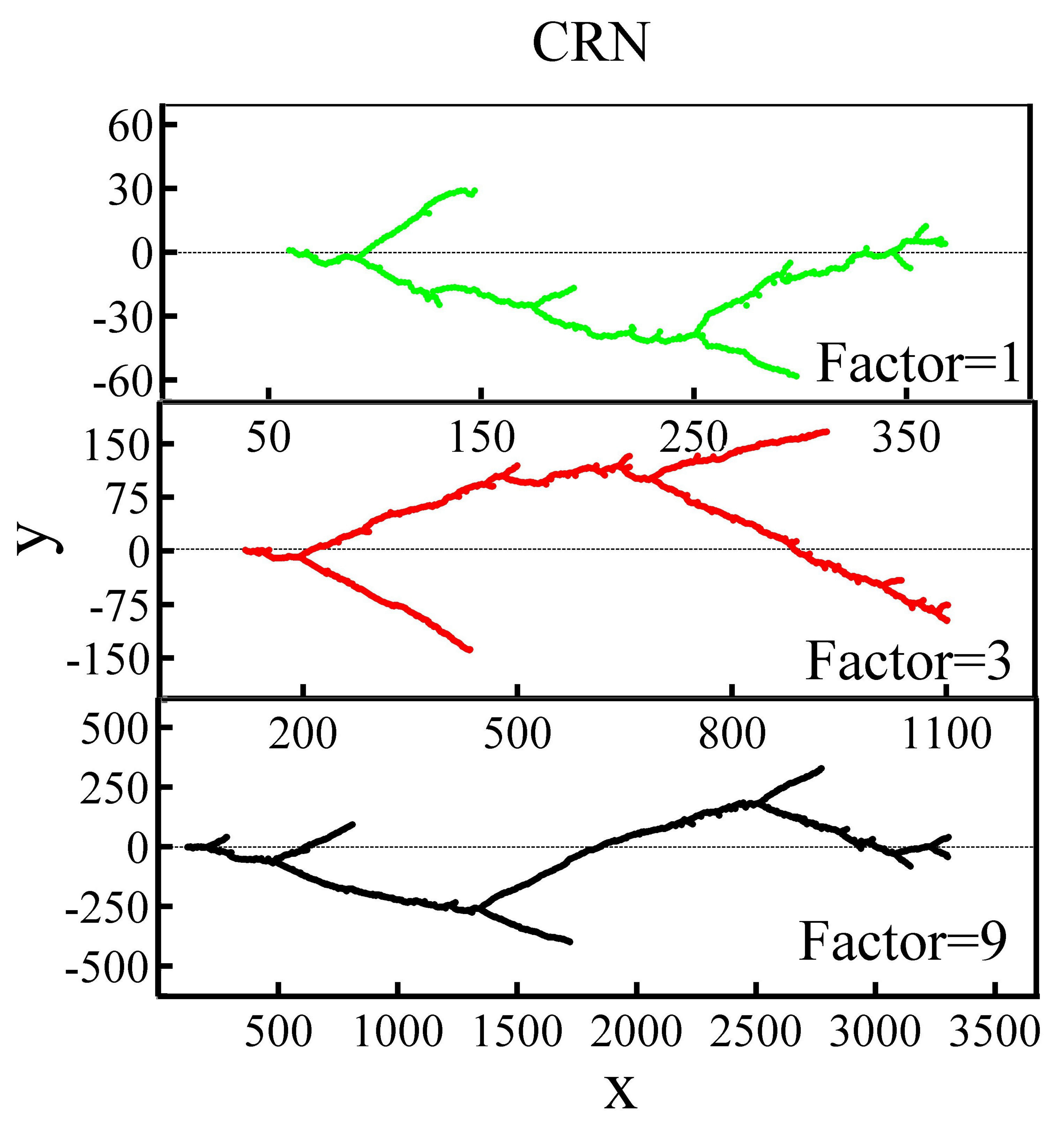}
(b)
\includegraphics*[width=7.5cm]{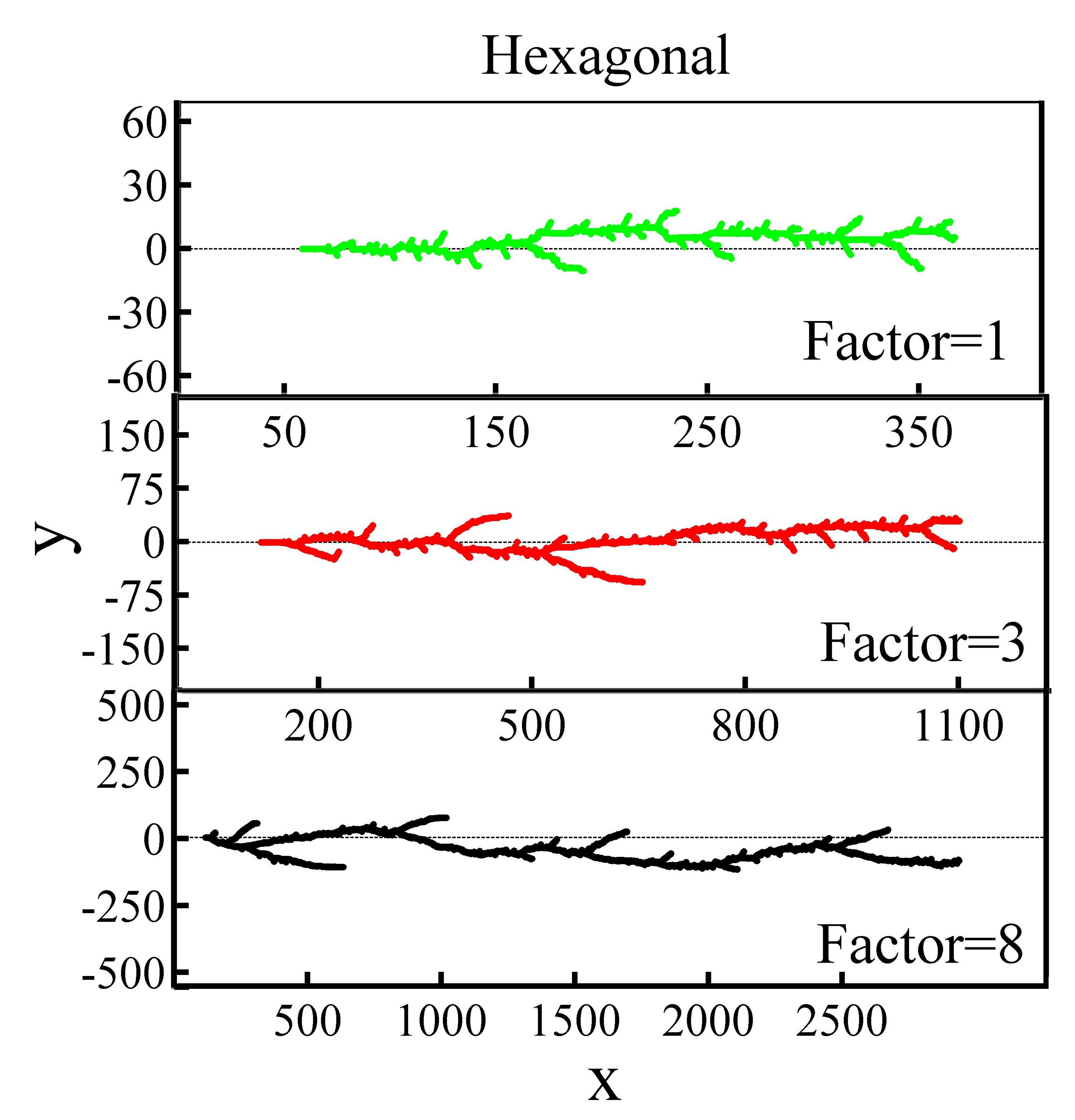}
}
\caption{(Color online) (a) The microbranching pattern using the continuous random network (CRN) model for $\Delta/\Delta_G=3.6$ and $\eta=2$ for $f=1$ in the upper curve,
for $f=3$ in the intermediate curve and for $f=9$ in the lower curve for continuous random network (CRN) model (b) The same for perturbed hexagonal lattice for $\Delta/\Delta_G=1.7$ and $\eta=0.25$.}
\label{pattarns2}
\end{figure}

In Figs. \ref{pattarns}(a) to \ref{pattarns}(c) we present the fracture pattern of the broken bonds for the small size perturbed honeycomb lattice
(that was used before in \cite{shay4},  called here $f=1$),
for the intermediate size lattice  ($f=3$), and for the large lattice ($f=9$), for three values of the driving displacement: small ($\Delta/\Delta_G=2.8$), intermediate ($\Delta/\Delta_G=3.4$),
and large ($\Delta/\Delta_G=4)$. The fracture patterns are plotted in the $x-y$ plane and depict the full simulated sample, when $x$ and $y$ have the units of $a_0$, the lattice scale.
In the large driving simulation for $f=1$ reported upon in our previous work, the microbranches reached the edge of the sample. In addition, in Fig. \ref{pattarns2}(a)
we present fracture patterns using the amorphous CRN model (that was used before in \cite{shay3}) with $\Delta/\Delta_G=3.6$ for the different sizes of lattices, and in Fig. \ref{pattarns2}(b), fracture
patterns using a  perturbed hexagonal lattice. 

In the Figures we can  immediately see the benefit of the larger
scale simulations; the noisy fracture patterns that were obtained using $f=1$ (the upper pattern in each figure), transform to the smoother and more physical-like patterns at $f=8-9$.
In Fig. \ref{pattarns2}(b) we can see clearly the curved
power-law shape of the microbranches (for a quantitative discussion, see Sec. \ref{results_c}).

\begin{figure}
\centering{
(a)
\includegraphics*[width=7.4cm]{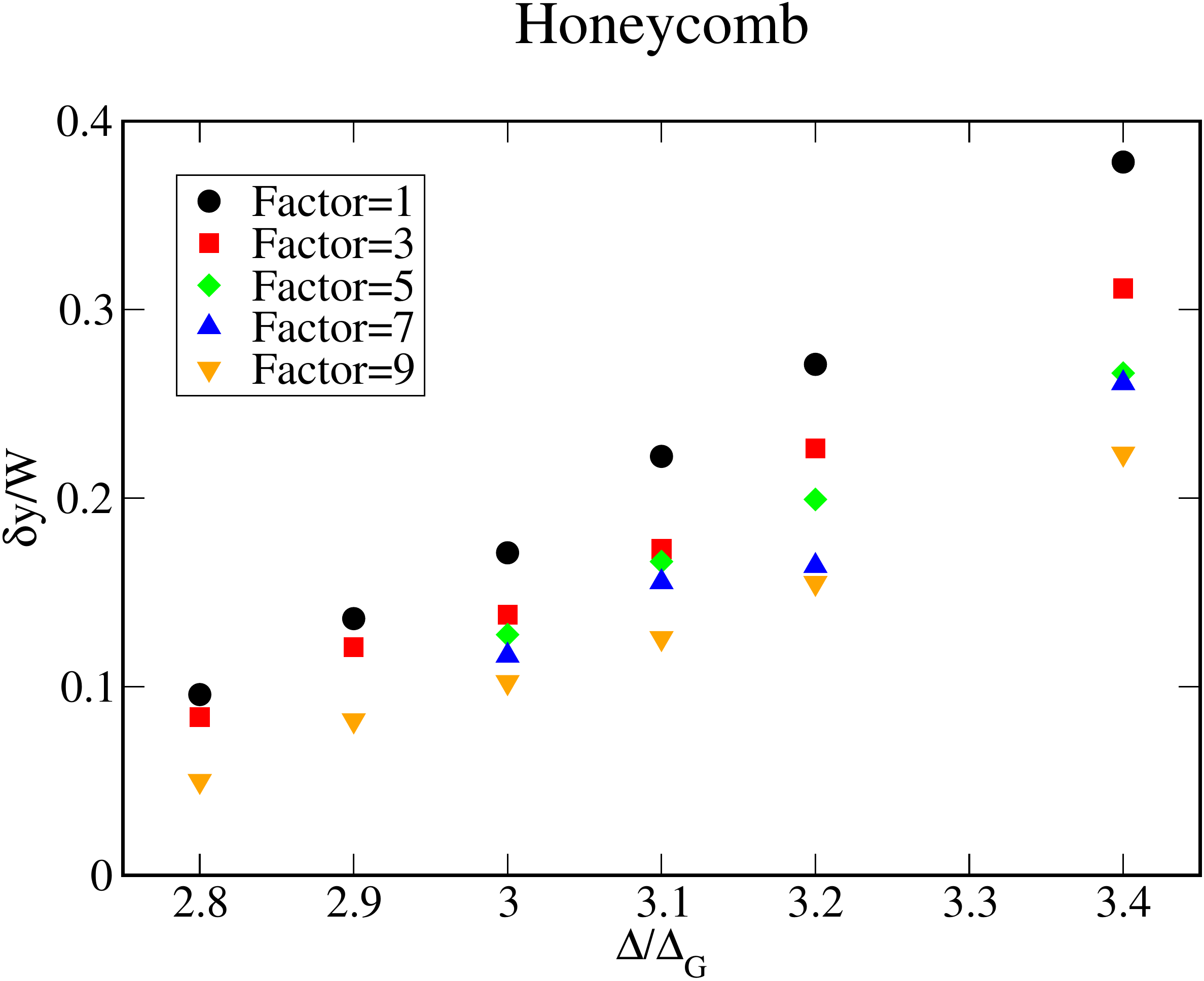}
(b)
\includegraphics*[width=7.4cm]{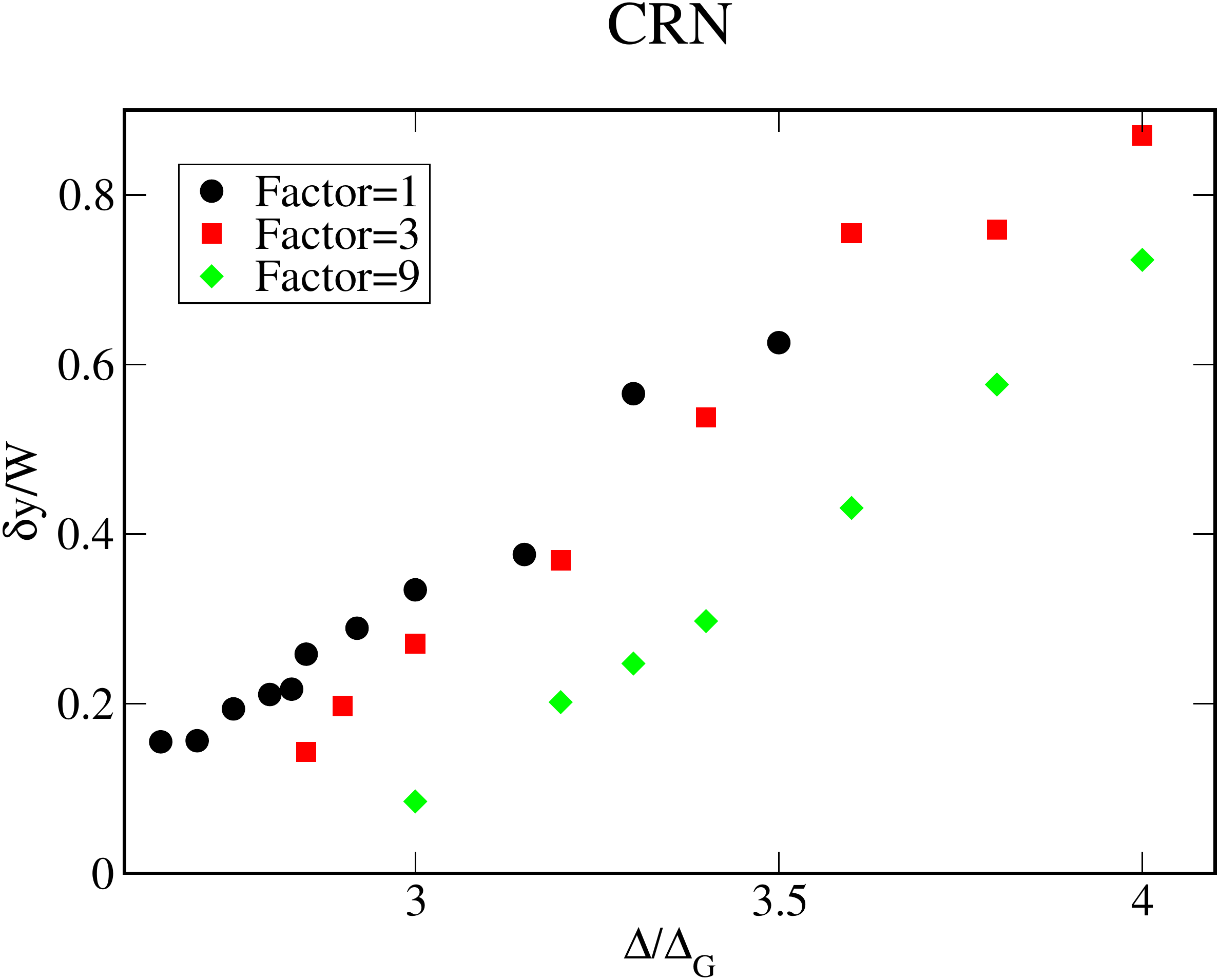}
(c)
\includegraphics*[width=7.4cm]{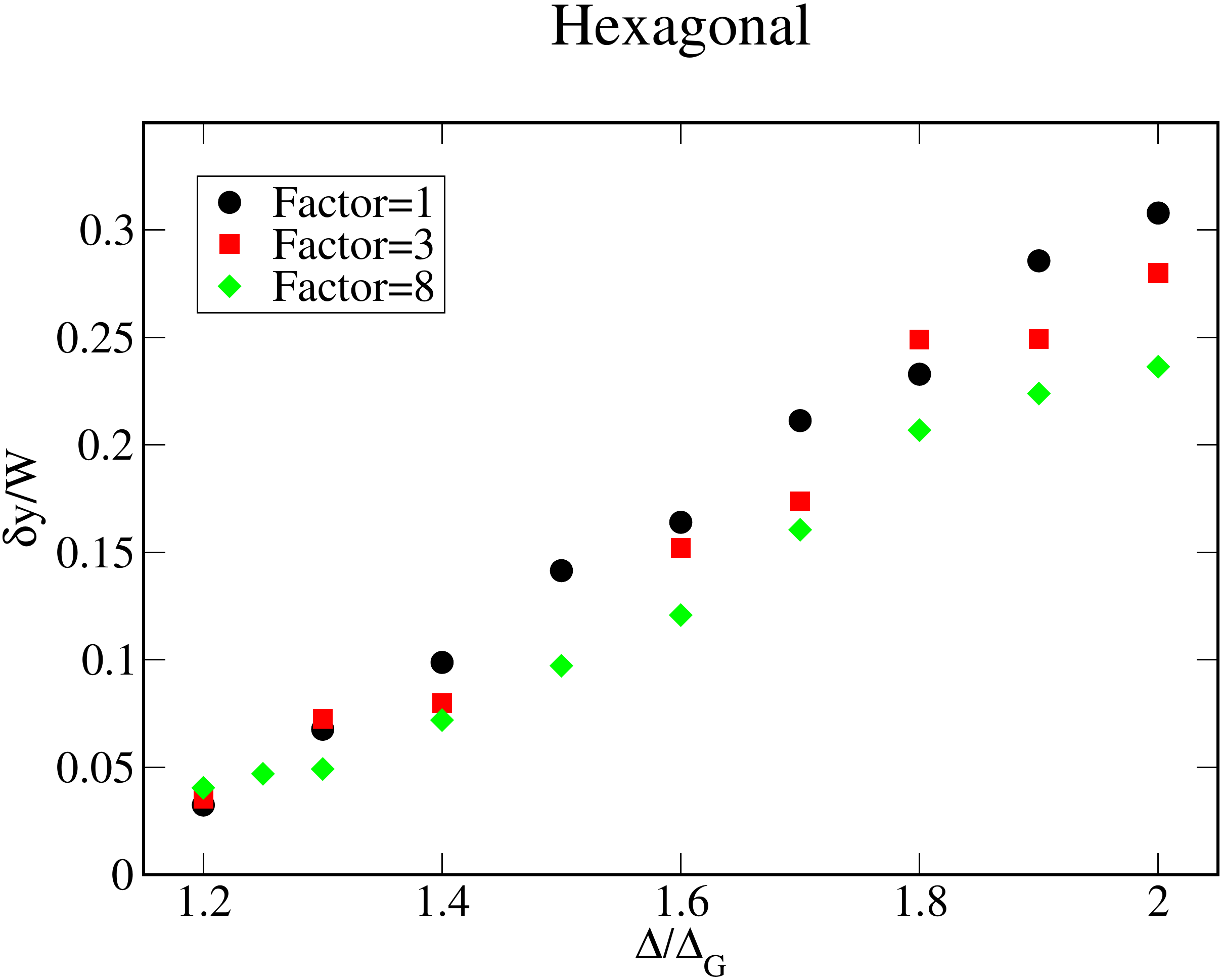}
}
\caption{(Color online) (a) The scaled width of the microbranching pattern $\nicefrac{\delta y}{W}$ for (a) a honeycomb perturbed lattice, (b) the CRN model  and (c) a hexagonal perturbed lattice, for different lattice sizes.  }
\label{deltay_w}
\end{figure}

A closer look reveals an important point: For a given driving displacement, the relative width of the fracture pattern decreases with the increase of the lattice size. 
This is crucial since otherwise the branching pattern in a macroscopic material would be macroscopic as well, against the evidence of
the experiments~\cite{fineberg_sharon2}.
In Fig. \ref{deltay_w}, we can see the relative fracture zone width $\nicefrac{\delta y}{W}$, when $\delta y$ is the width of the microbranching
pattern (see Fig. \ref{pattarns}(b) for a pictorial explanation), defined as the difference between the maximum and minimum $y$'s of broken bonds,  and $W$ is the sample width. For any given value of $\nicefrac{\Delta}{\Delta_G}$ the normalized width of the microbranching pattern decreases with the lattice width. This effect in seen clearly  in the perturbed honeycomb
lattice and in the CRN lattice, and in a more moderate way in the hexagonal perturbed lattice. This result is crucial; if the lattice models are physical, then when increasing
the lattice size, the relative fracture zone  must decrease, so that the branching does not remain macroscopic in the
$N\to\infty$ limit, which would conflict with the experimental results). It can be seen that the dependence of $\delta y$ on $\Delta$ is more or less linear. This  may be
related to the linear increase of the size of the average microbranch as a function of the driving displacement (for example, see  Fig. \ref{hex_res}(b))below).

\section{Results}
\label{results}

\subsection{The crack's tip shape}
 
As mentioned in the introduction, the new experiments using gels yield clear snapshots of the
crack tip~\cite{gels2,livne,review_new}. Larger scale simulation enable us to compare the crack tip shape to the experimental snapshots in both steady-state cracks and
near the origin of instability.

\begin{figure}
\centering{
\includegraphics*[width=16cm]{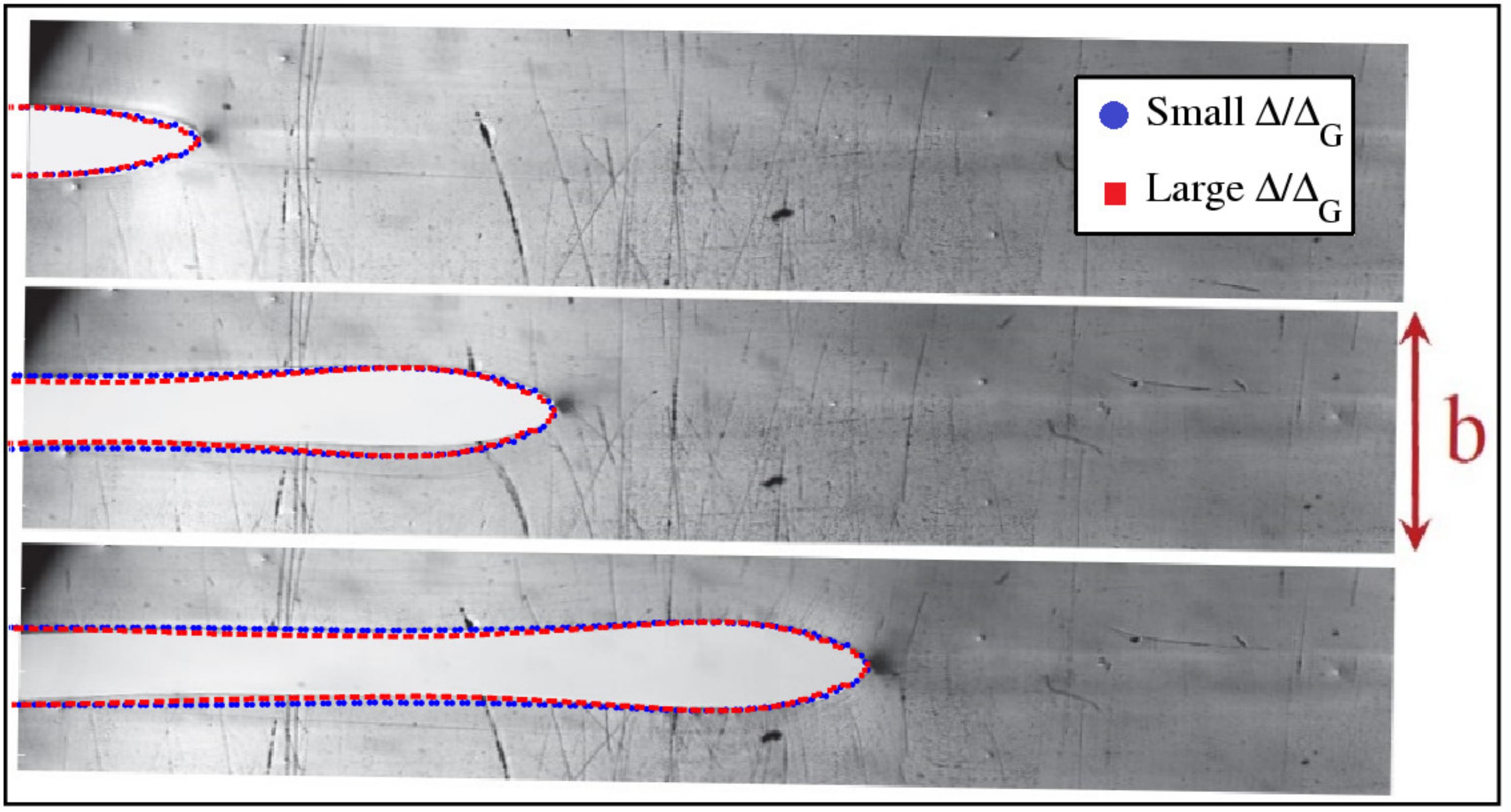}
}
\caption{(Color online) Several snapshots  of the crack's tip in steady-state mode-I fracture in experiments on gels. In the upper picture, when the crack length is small comparing
to the sample's width, the crack has the (well-known) parabolic shape. As the crack length grows, the finite size of the strip effects the crack's tip shape, generating a ``tadpole-like"
shape. The lattice simulations for the crack tip shape are added to the snapshots in the dashed curves. The experimental pictures are taken from~\cite{review_new}.}
\label{tadpole}
\end{figure}

First, we present the very good agreement between experiment and simulations of the crack's tip shape in steady-state cracks in the finite-width strip geometry in Fig \ref{tadpole}.
In the upper picture, when the crack length is small comparing
to the sample's width, the crack has the (well-known) parabolic shape; no blunting can be seen and LEFM
works perfectly (except for nonlinear effects in the extreme crack tip zone~\cite{livne}). As the crack length grows, the finite width of the strip affects
the crack's tip shape, generating a ``tadpole-like" shape. We set a careful lattice simulations in finite width strip using different values of $\Delta/\Delta_G$.
Since the simulations have only a finite number of atoms, the results are
scaled to the real size of the experimental sample. We can see in Fig. \ref{tadpole} the excellent match of the crack's tip shape between
pure lattice simulations (described in detail below) and the experiments. This ``tadpole" shape is generic for finite-size (strip)
lattice simulations, both honeycomb or hexagonal, with any amount of viscosity. Since the experimental crack is not exactly symmetric between the
upper and the lower size of the crack, we used several values of $\Delta/\Delta_G$ to get an optimal match between experiment and simulation (the small $\Delta/\Delta_G$ shape is somewhat
better on the upper side and the large $\Delta/\Delta_G$ shape is somewhat better in the lower side).

\begin{figure}
\centering{
\includegraphics*[width=8.5cm]{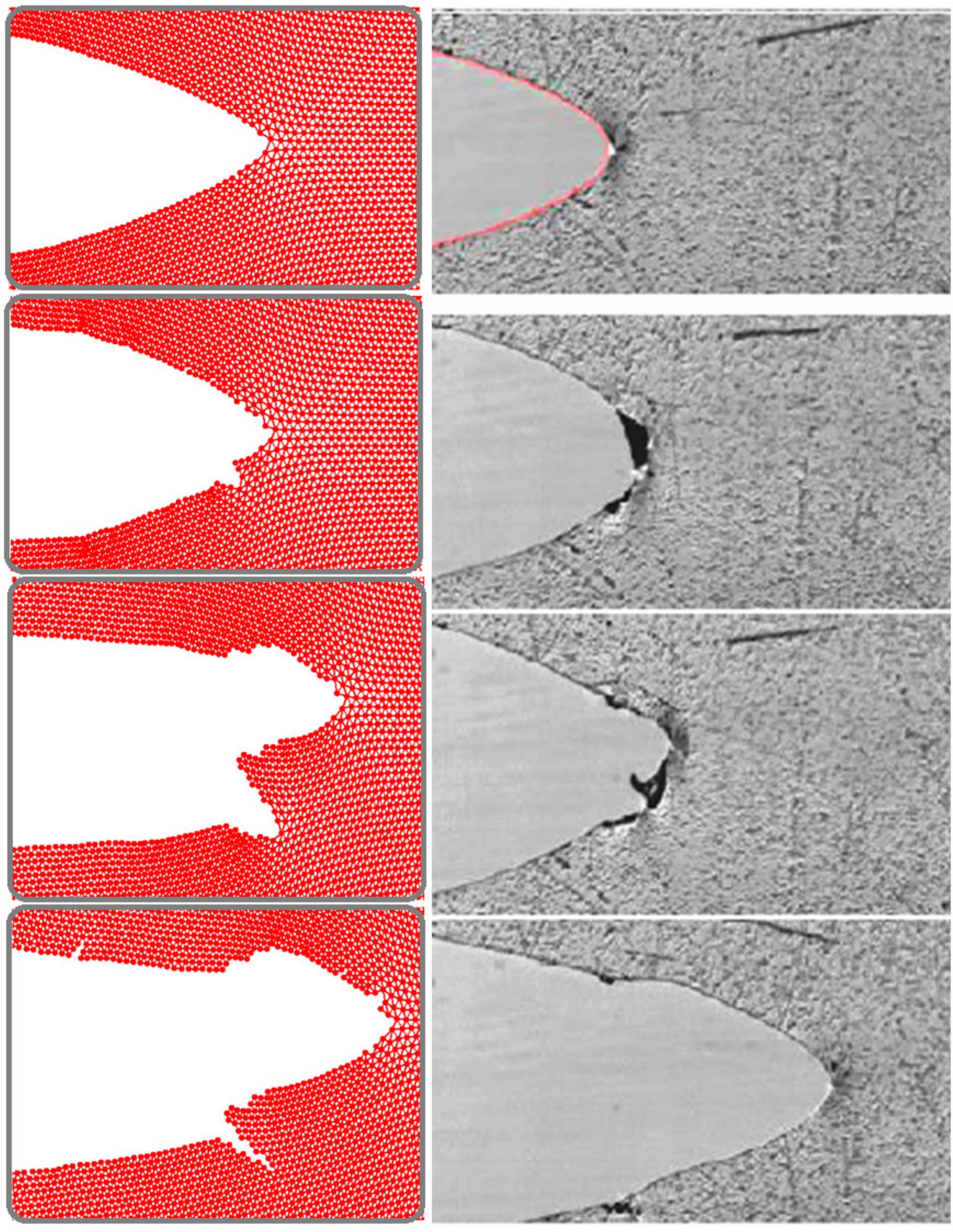}
}
\caption{(Color online) Several snapshots of the crack tip near the origin of the microbranching instability. on the right, there is a series of experimental snapshots of the crack
tip before, during, and right after a new microbranch is born. On the left, there is a series of snapshots of our lattice simulations. Qualitatively, there is a very close resemblance
between the two series. The experimental pictures are taken from~\cite{review_new}.}
\label{origin_micro}
\end{figure}

Moreover, in ~\cite{review_new} there are snapshots of the crack tip shape near the origin of the microbranching instability (on the right in Fig. \ref{origin_micro}). In order to reproduce
this crack tip shape (in addition to using the large scale simulation), we set $\varepsilon=2a_0$ (only for this part in the paper), to magnify the crack tip radius compared to the lattice
scale $a_0$. This enables us to meaningfully compare the simulation results to experiments (this value of $\varepsilon$ is a bit extreme, but qualitatively, the same physical effect can be seen using smaller
value of $\varepsilon$). The lattice simulations are presented on the left in Fig. \ref{origin_micro}).

We can see that there is a close resemblance between the two series's of snapshots. At first, the crack travels in the midline of the sample, yielding the parabolic crack's tip
shape of LEFM. When a new microbranch is born, the crack tip bifurcates to two, while very rapidly one ``wins" and continues to propagate, and the other ``dies". The branch
that continues to propagate has a small deformed shape comparing to the original shape before it continue to propagate (the crack shape here significantly differs from the LEFM prediction).
All of these features are reproduced in the lattice simulations. It is important to mention that both experiments and simulations reveal that the origin of
microbranching in mode-I fracture  always lies in the immediate vicinity of the crack tip itself, rather than far behind the main crack.

\subsection{Length of microbranches}

\begin{figure}
\centering{
(a)
\includegraphics*[width=7cm]{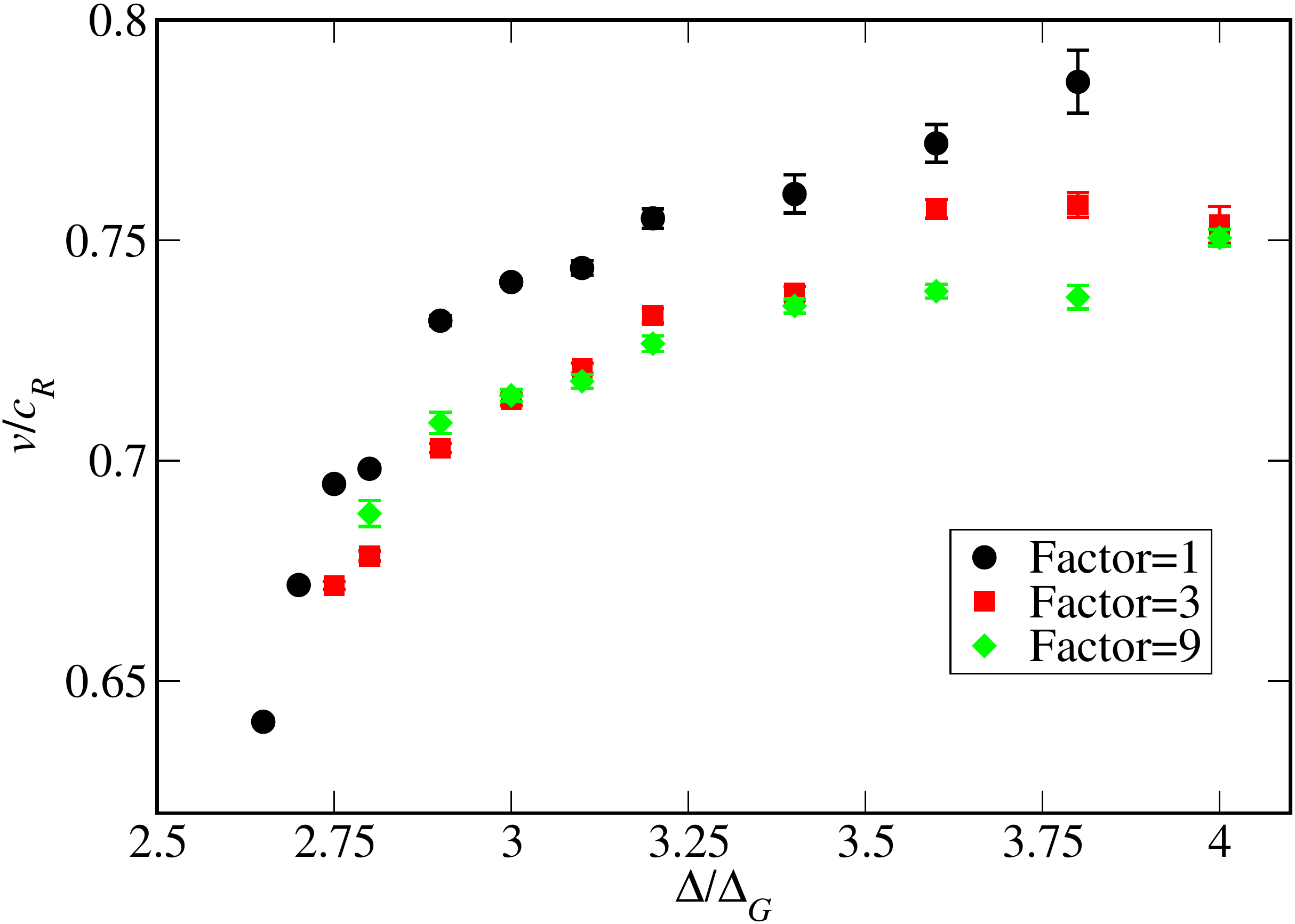}
(b)
\includegraphics*[width=7cm]{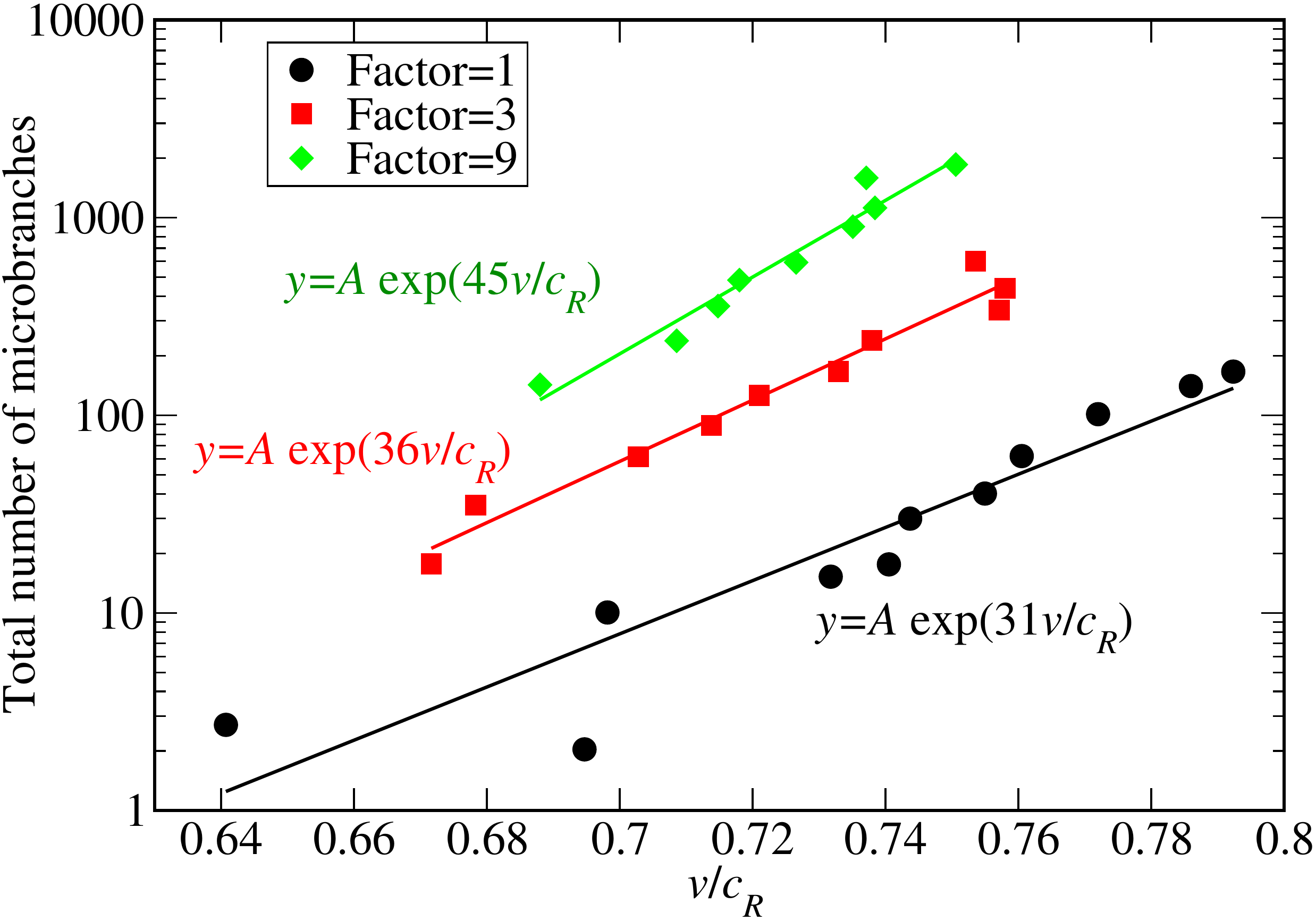} \\
(c)
\includegraphics*[width=7cm]{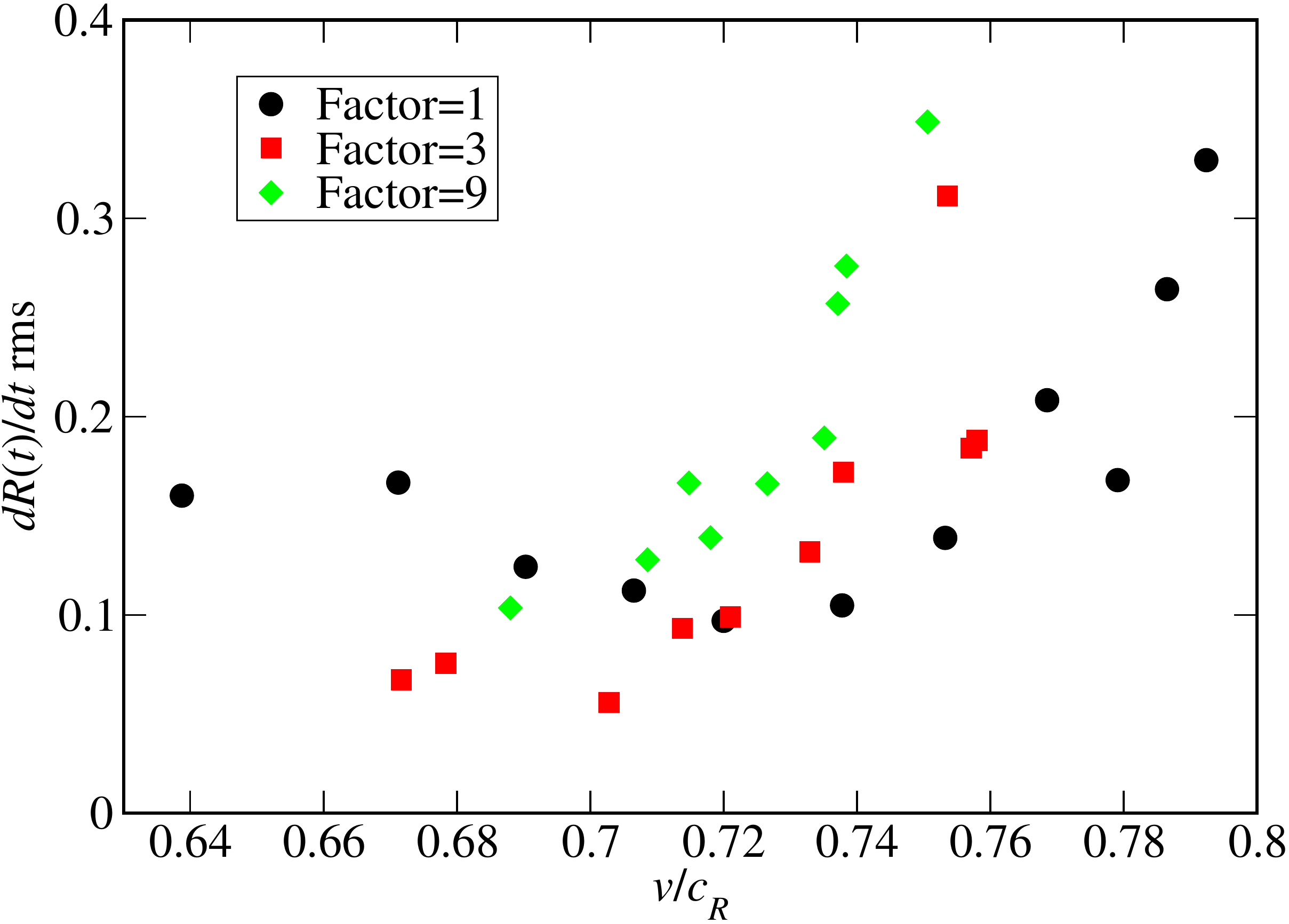}
}
\caption{(Color online) (a) The $v(\Delta)$ curve of a perturbed honeycomb lattice using different lattice sizes. (b) Total number of broken bonds in the microbranches
as a function of the crack velocity. (c) The RMS of the derivative of electrical resistance with respect to the time as a function of the crack velocity.}
\label{hon_res}
\end{figure}

\begin{figure}
\centering{
(a)
\includegraphics*[width=7cm]{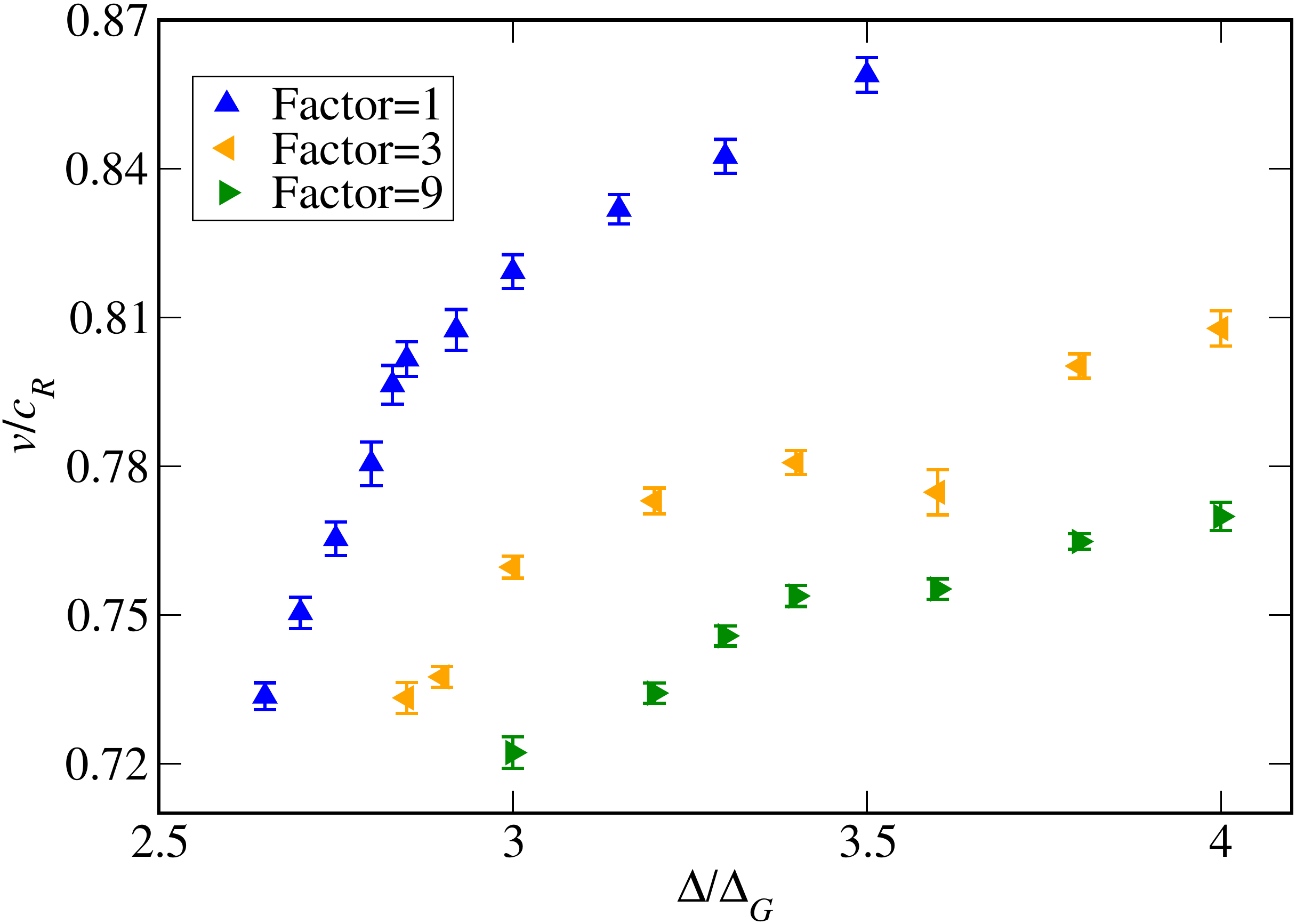}
(b)
\includegraphics*[width=7cm]{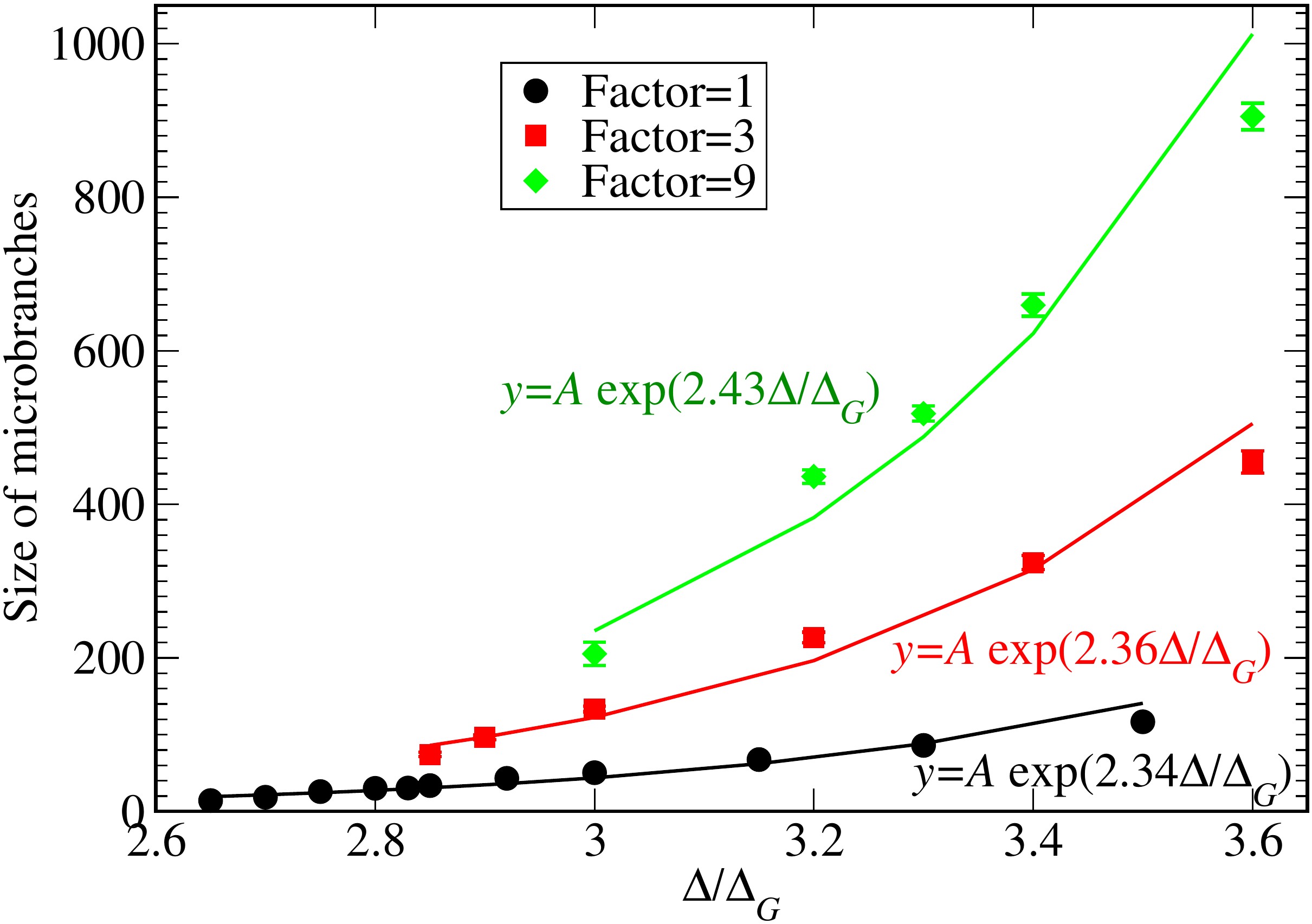}
}
\caption{(Color online) (a) The $v(\Delta)$ curve of a CRN lattice using different lattice sizes. (b) Total amount of broken bonds in microbranches.
as a function of the crack velocity.}
\label{crn_res}
\end{figure}

In this section we present the quantitative results for various features of the fracture patterns, especially  the microbranches.
First, we present the $v(\Delta)$ curve, the total amount of broken bonds in the microbranches, measured for a crack that
reached the end of the strip, as a function of the crack velocity and the RMS of $dR(t)/dt$, the rate of increase of the derivative of the electrical
resistance with respect to time as a function of the crack velocity. The results are normalized to to $c_R$, the Rayleigh wave speed, which is calculated for
different values of $k_{\theta}$ in Appendix \ref{rayleigh}. The results for the perturbed honeycomb lattice is presented in Fig. \ref{hon_res}, and for the CRN in Fig. \ref{crn_res}.

We can clearly see that in all three models the slope of the $v(\Delta)$ appears to saturate in the high velocity regime for the larger system sizes. This saturation is known from previous 
lattice studies~\cite{shay1}. The shape of the curves are similar to the experimental $v(\Delta)$ curves of real amorphous materials~\cite{review}.
A close look at the curves of the total amount of broken bonds in the microbranches (which is our proxy for the average length of a microbranch as measured experimentally,
which we use to reduce the statistical noise)
as a function of the crack velocity, using both the honeycomb perturbed lattice and the CRN, reveals
quantitatively what we have seen qualitatively using small system sizes. Due to the noisiness of using finite size lattices, rather than two different regimes, with a single steady state crack
 at small velocities,
and a sharp increase in the length of a microbranch in the large velocity regime, we get a smooth exponential behavior. However, the exponent
of the curves increases robustly with the lattice size, sharpening the difference between the steady state  regime and the the microbranching (large velocity) regime. In addition
we can see that in general, the CRN results act like a very perturbed honeycomb lattice, i.e., in any given $\Delta/\Delta_G$, the number of broken bonds is larger in the CRN than in
the perturbed honeycomb lattice, similar to what we saw in Ref.~\cite{shay4}. In addition, using the honeycomb perturbed lattice we can see the increase of the RMS amplitude
of the derivative of the electrical resistance with respect to the time as a function of the crack velocity. At high velocities, the RMS amplitude is approximately three times
greater than the amplitude in low crack velocities, in agreement with the experiments regarding the RMS of the crack velocity (Fig. 11(c) in~\cite{fineberg_sharon2}).

\begin{figure}
\centering{
(a)
\includegraphics*[width=7cm]{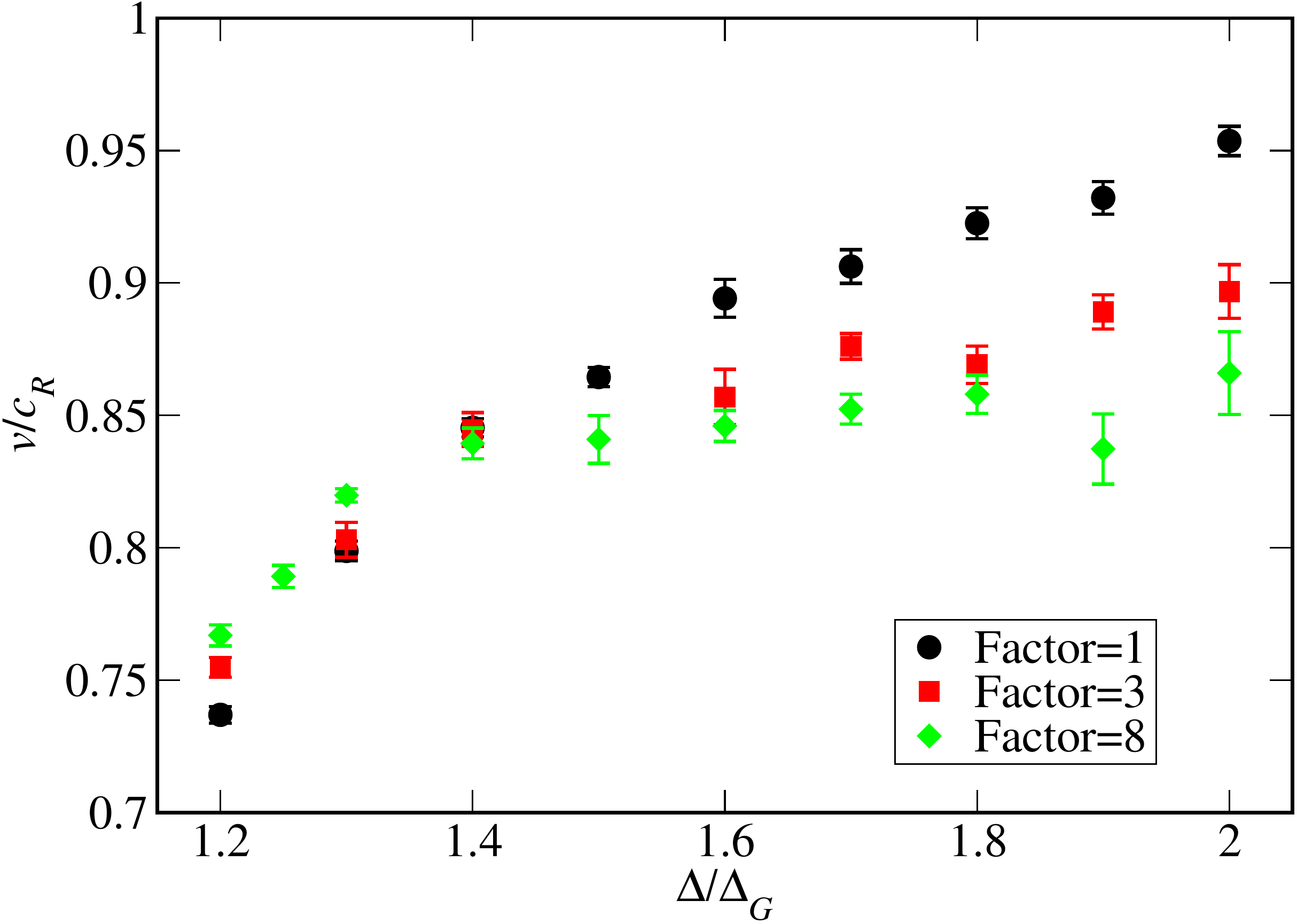}
(b)
\includegraphics*[width=7cm]{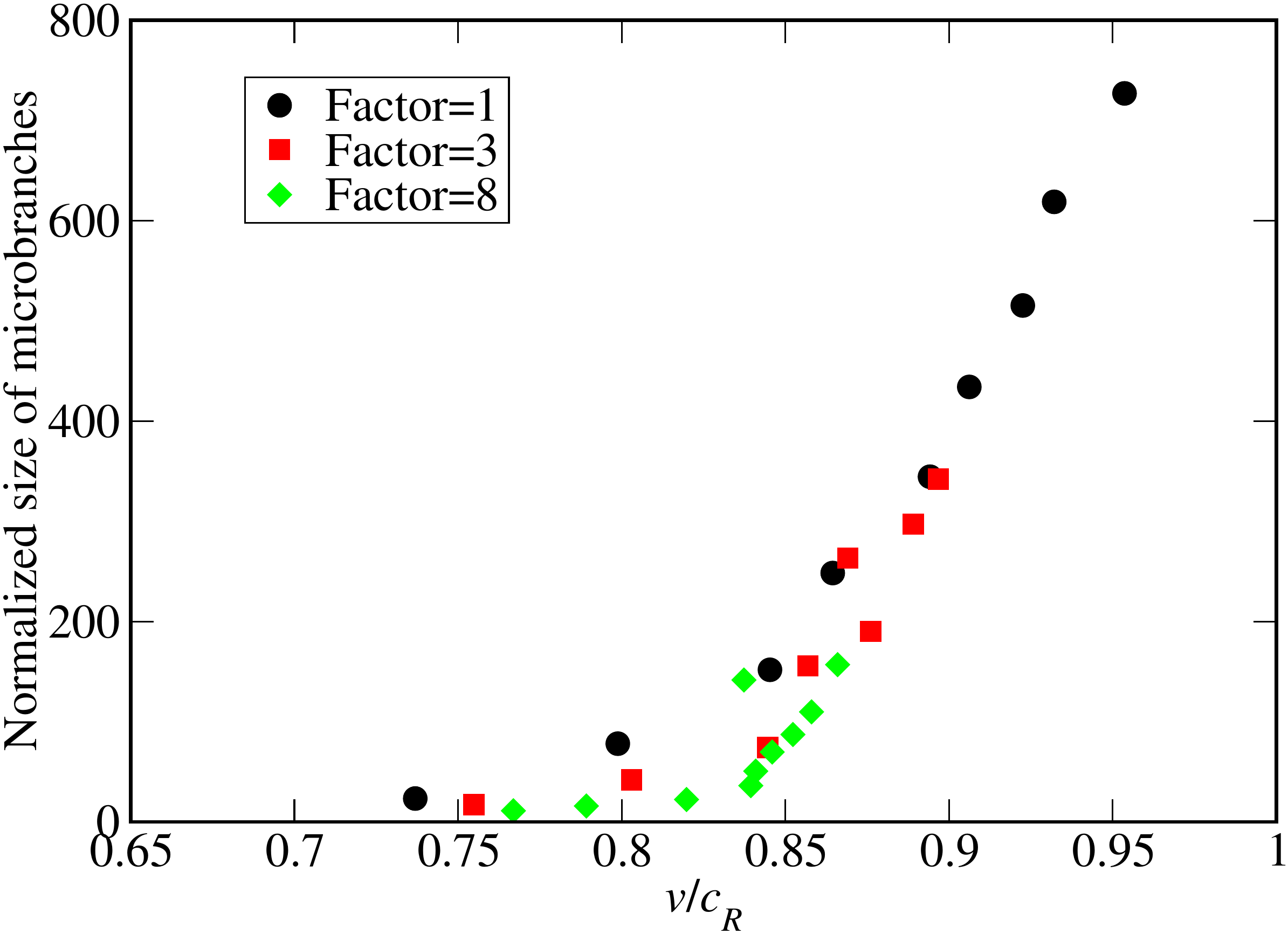}
}
\caption{(Color online) (a) The $v(\Delta)$ curve of a perturbed hexagonal lattice using different lattice sizes. (b) Total size of broken bonds in microbranches.
as a function of the crack velocity.}
\label{hex_res}
\end{figure}
Looking at the results for the perturbed hexagonal lattice, we get one of the most important results of this paper. In Fig. \ref{hex_res}(b), we present the
normalized number of microbranches broken bonds
(per number of bonds in the entire system) as a function of the crack velocity. The different curves (using various lattice size) are similar to each other, however, enlarging the lattice size,
the transition become sharper with $f$. Using $f=8$, we can see two different separate regimes (one of steady-state cracks, and one for microbranching), that
looks very much like the transition of the average microbranch length in the experiments (Fig. 11(a) in~\cite{fineberg_sharon2}). This result verifies the main assumption of the lattice models.
The physical phenomena of microbranching is qualitatively described by lattice models and simulations, when enlarging the system size, the results become more quantitatively similar to the
(macroscopic) experimental results.

\begin{figure}
\centering{
(a)
\includegraphics*[width=7.5cm]{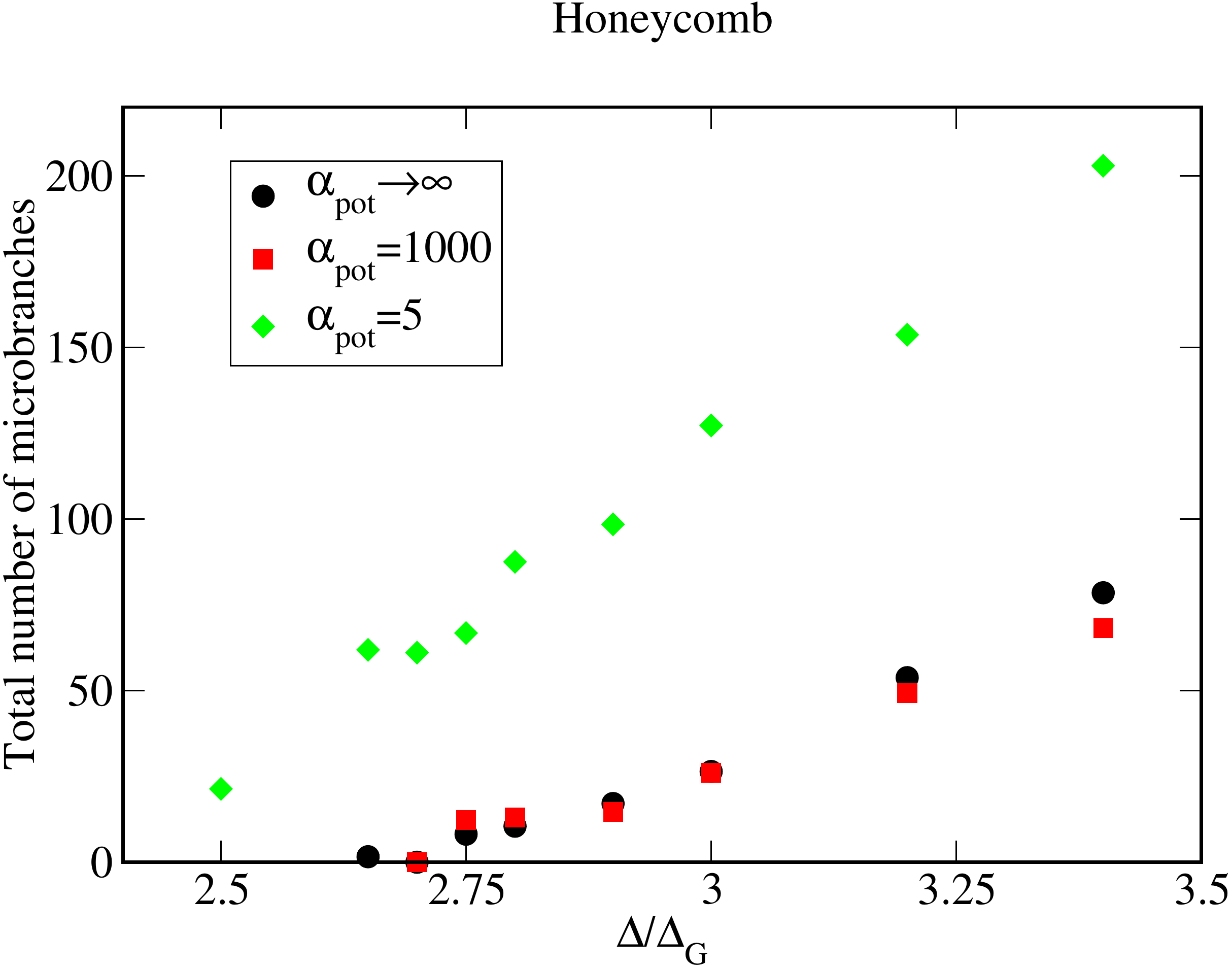}
(b)
\includegraphics*[width=7.5cm]{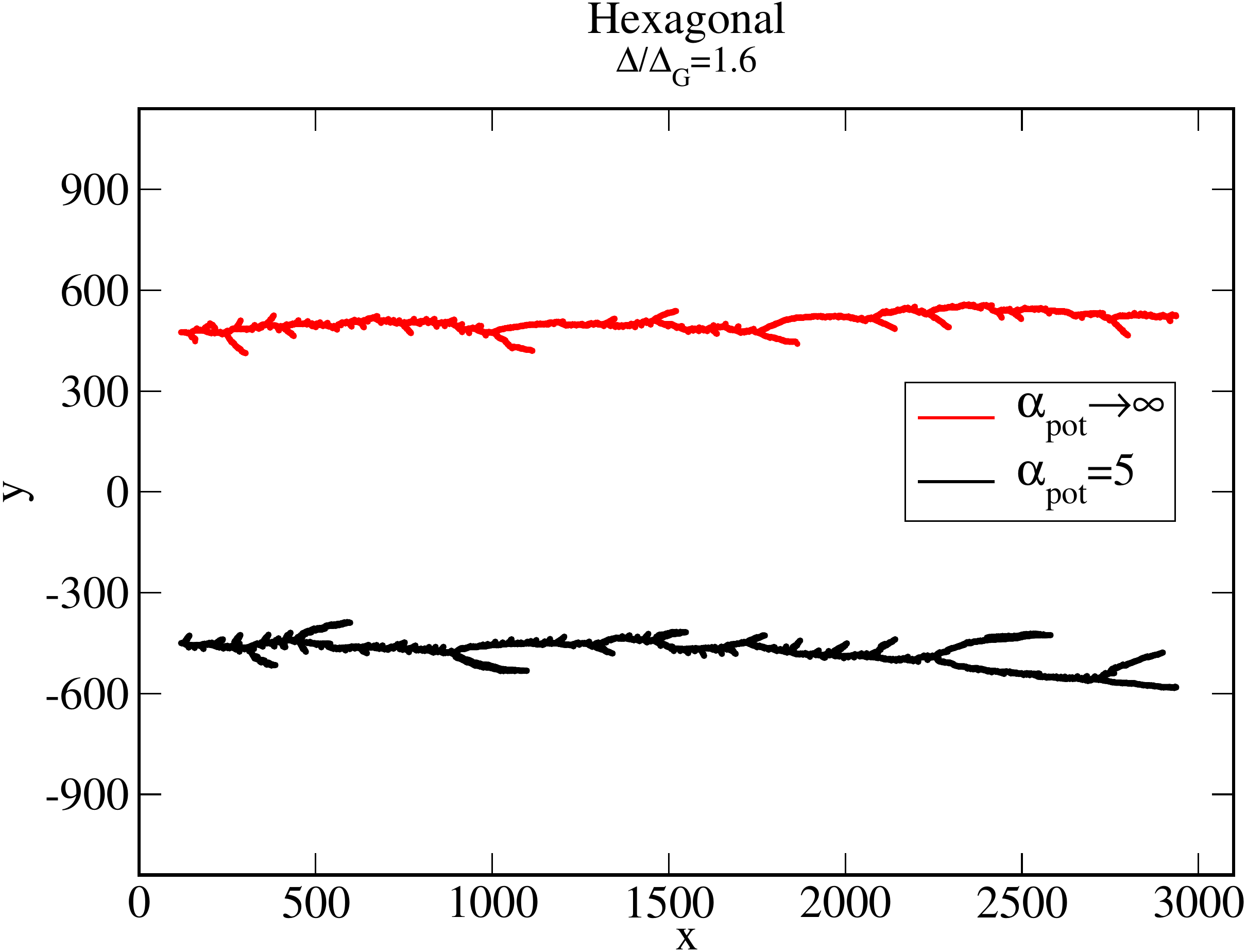}
}
\caption{(Color online) (a) The total size of microbranches as a function of the driving displacement $\Delta/\Delta_G$ using different values of $\alpha_{\mathrm{pot}}$ along with
the piecewise linear model ($\alpha_{\mathrm{pot}}\to\infty$) using perturbed honeycomb lattice with $f=1$. (b) The microbranching pattern in a perturbed hexagonal lattice
using larger scale lattices ($f=8$) for $\Delta/\Delta_G=1.6$
and $\eta=0.25$ for $\alpha_{\mathrm{pot}}\to\infty$ in the upper curve, for $\alpha_{\mathrm{pot}}=5$ in the lower curve.}
\label{nonlinear_fig}
\end{figure}

Nevertheless, quantitatively, the critical velocity that was found in the experiments was less than the critical velocities that were seen in our simulations. However, in previous studies
it was shown that in the lattice models, we can control the critical velocity using different values of $\alpha_{\mathrm{pot}}$~\cite{shay1,shay2}. Here we check that this is still valid when
using a 3-body force-law and a  perturbed lattice. In Fig. \ref{nonlinear_fig}(a), for the case $f=1$, we present the total amount of microbranches using the perturbed honeycomb lattice. We can see that first,
$\alpha_{\mathrm{pot}}=1000$ reproduces the piecewise linear results, as expected. Second,
smaller values of $\alpha_{\mathrm{pot}}$ yields a much larger quantity of broken bonds for a given $\Delta$, indicating a much lower critical velocity (because of the noise in this lattice 
we cannot realize the zero microbranch regime). In addition we check whether using larger-scale lattices ($f=8$) yields the same fracture patterns. In Fig. \ref{nonlinear_fig}(b) we can see
that fracture patterns remain similar using the smaller value of $\alpha_{\mathrm{pot}}$ (here we used the perturbed hexagonal lattice), though for a given $\Delta$ the microbranches
are larger, as expected.

\subsection{Microbranching statistics}
\label{results_c}

Having larger systems enables, for the first time, the generation of enough statistics to examine important quantitative features
of the microbranches that have been measured experimentally~\cite{fineberg_sharon2,review}. Since in this study, we obviously are restricted to 2D features only, we focused here on
the branching angle of the microbranches, and on the power-law shape of the microbranches.

\begin{figure}
\centering{
(a)
\includegraphics*[width=7cm]{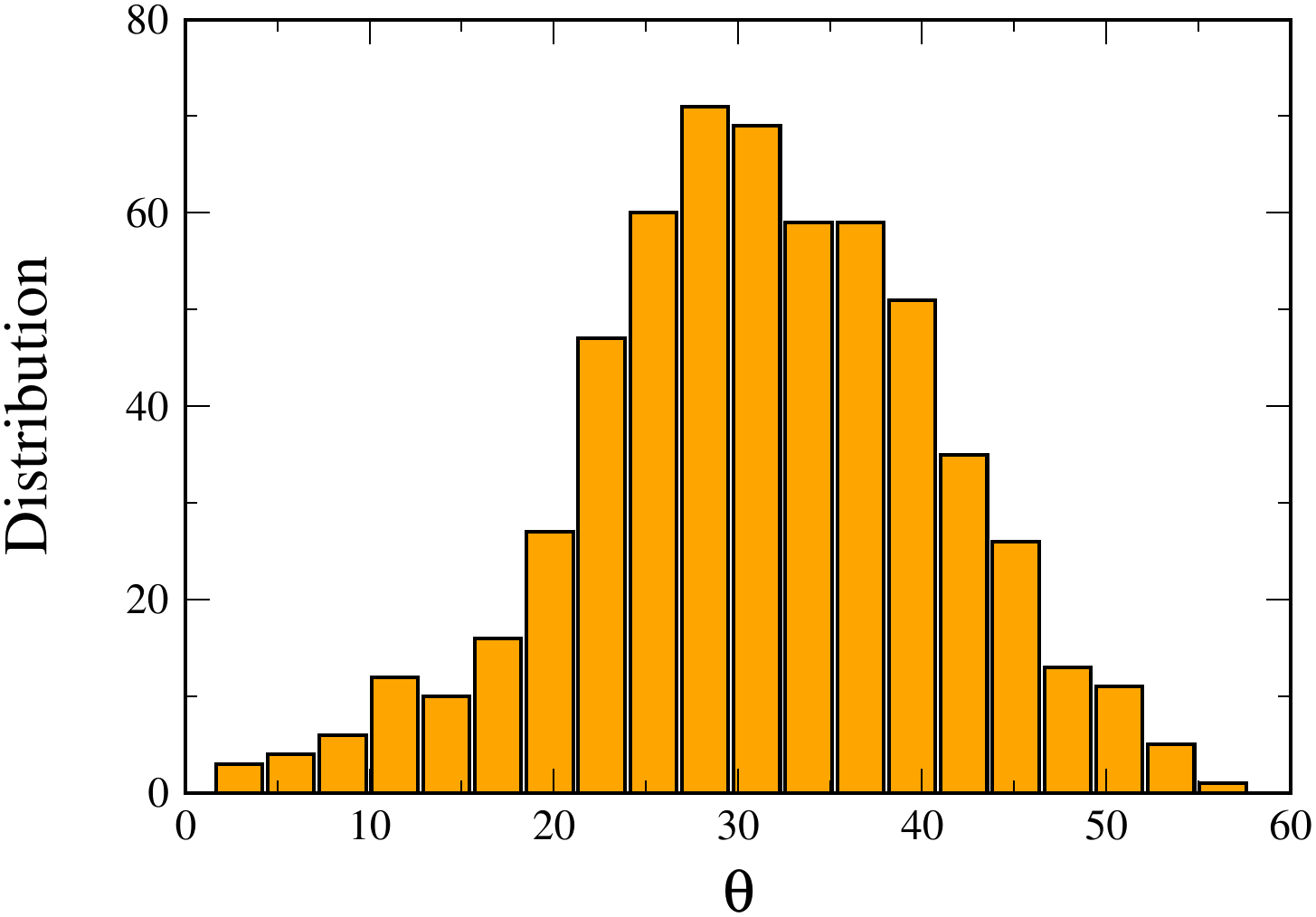}
(b)
\includegraphics*[width=7cm]{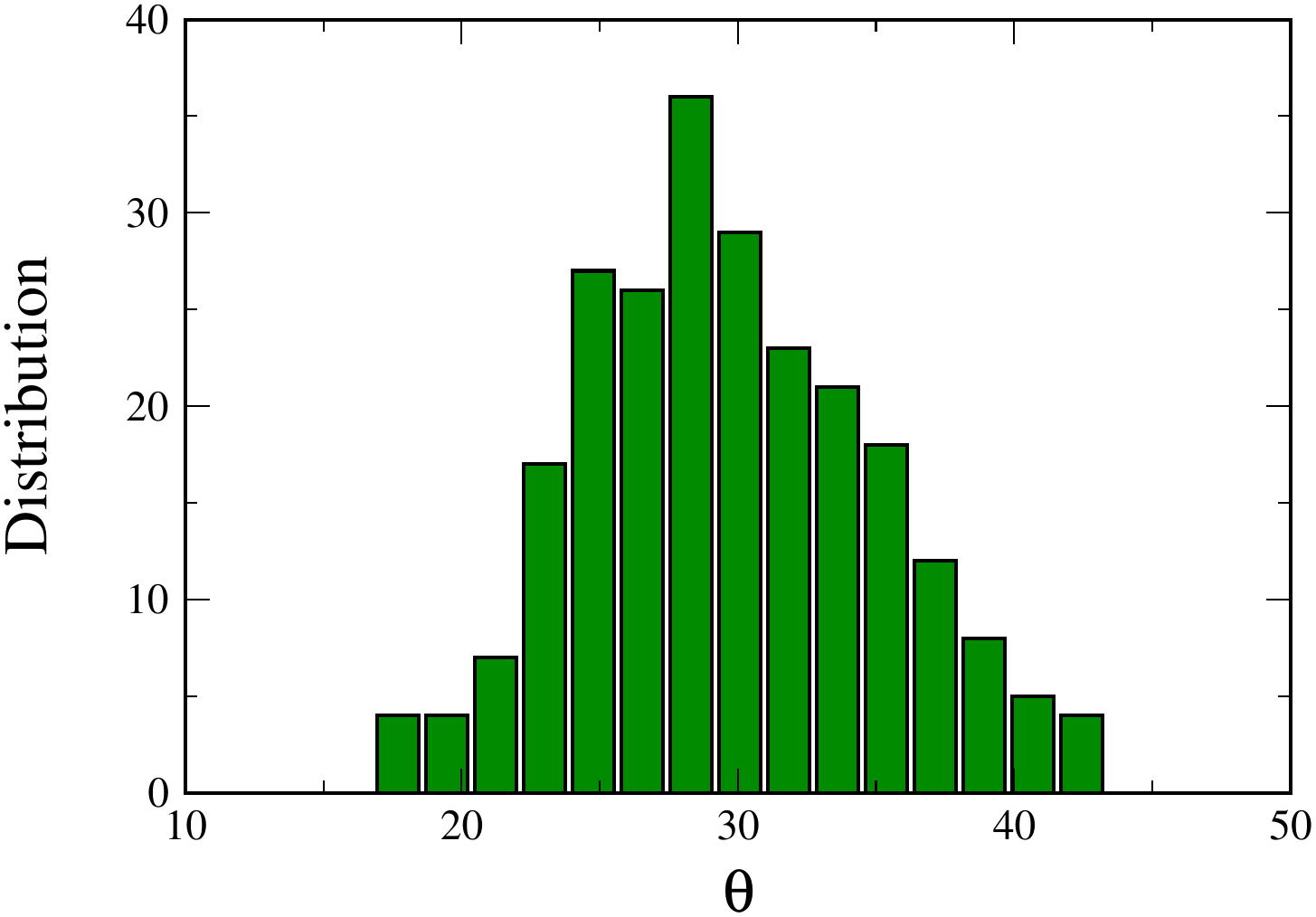} \\

\vspace*{0.1in}
(c)
\includegraphics*[width=7cm]{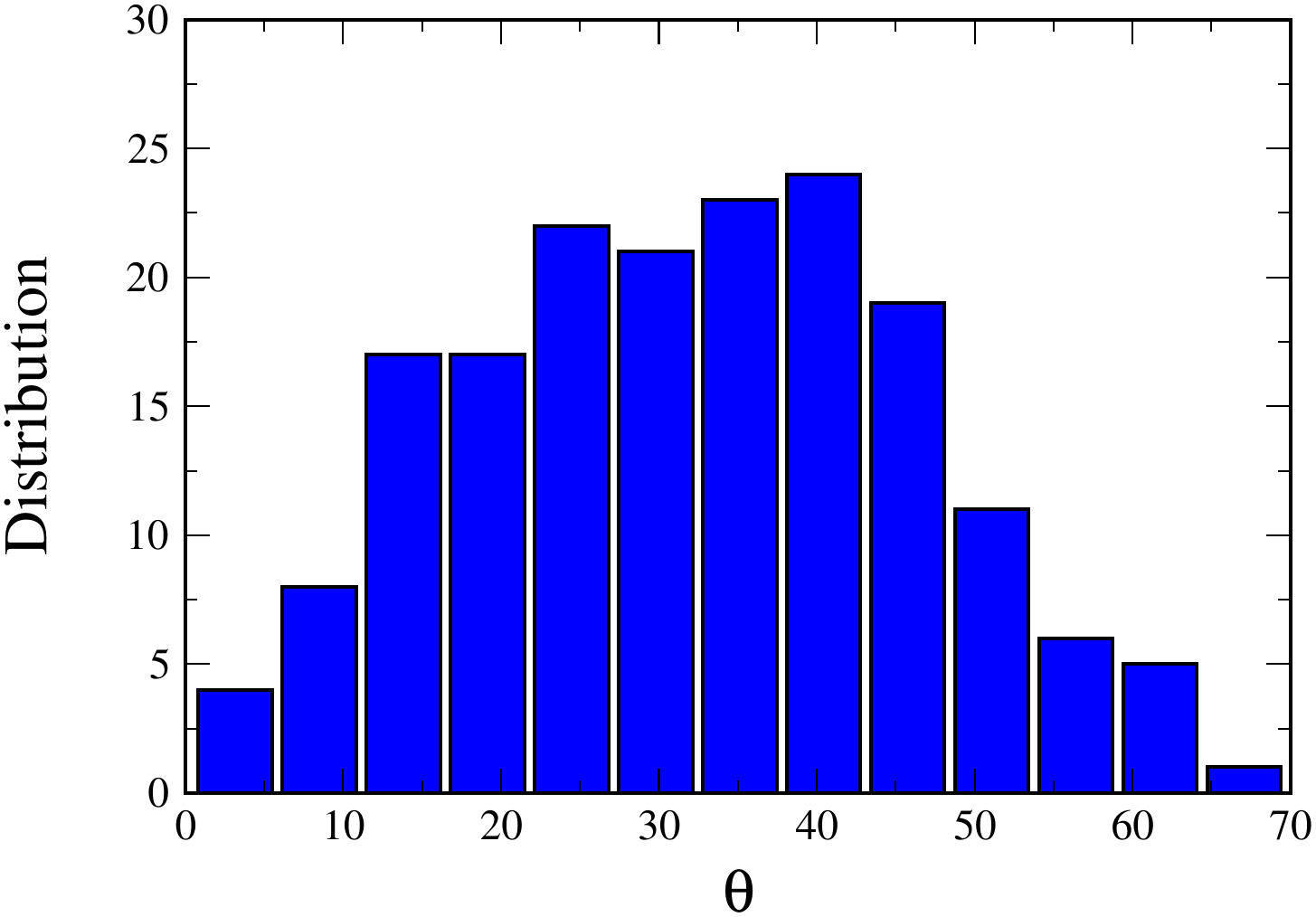}
}
\caption{(Color online) The distribution of the branching angle of the microbranches arise nearby the main crack using (a) a perturbed honeycomb lattice, (b) the CRN model  and (c) a perturbed hexagonal lattice.}
\label{hist_teta}
\end{figure}

The experimental studies on PMMA finds a narrow distribution of the branching angle between $20\degree\leqslant\theta\leqslant40\degree$, with an average
angle of $30\degree$~\cite{fineberg_sharon2}. We note that a previous study using a random perturbed Born-Maxwell model~\cite{bm2} yielded the wrong branching angle, namely $15\degree$. 
In Fig. \ref{hist_teta} we present the branching angle distribution of all the microbranches generated using all the values of $\Delta/\Delta_G$ in
the different models that were used in this study. We can see that in all the models studied, the average branching angle is near $30\degree$, very much like the experimental
branching angle. The variance is different using the different models, where the variance of the CRN lattice is the narrowest, and thus most similar to the experiments. We also note that
this result corresponds nicely with the LEFM microbranch analysis of~\cite{katzav} which yields a $27\degree$ branching angle. 

\begin{figure}
\centering{
(a)
\includegraphics*[width=7cm]{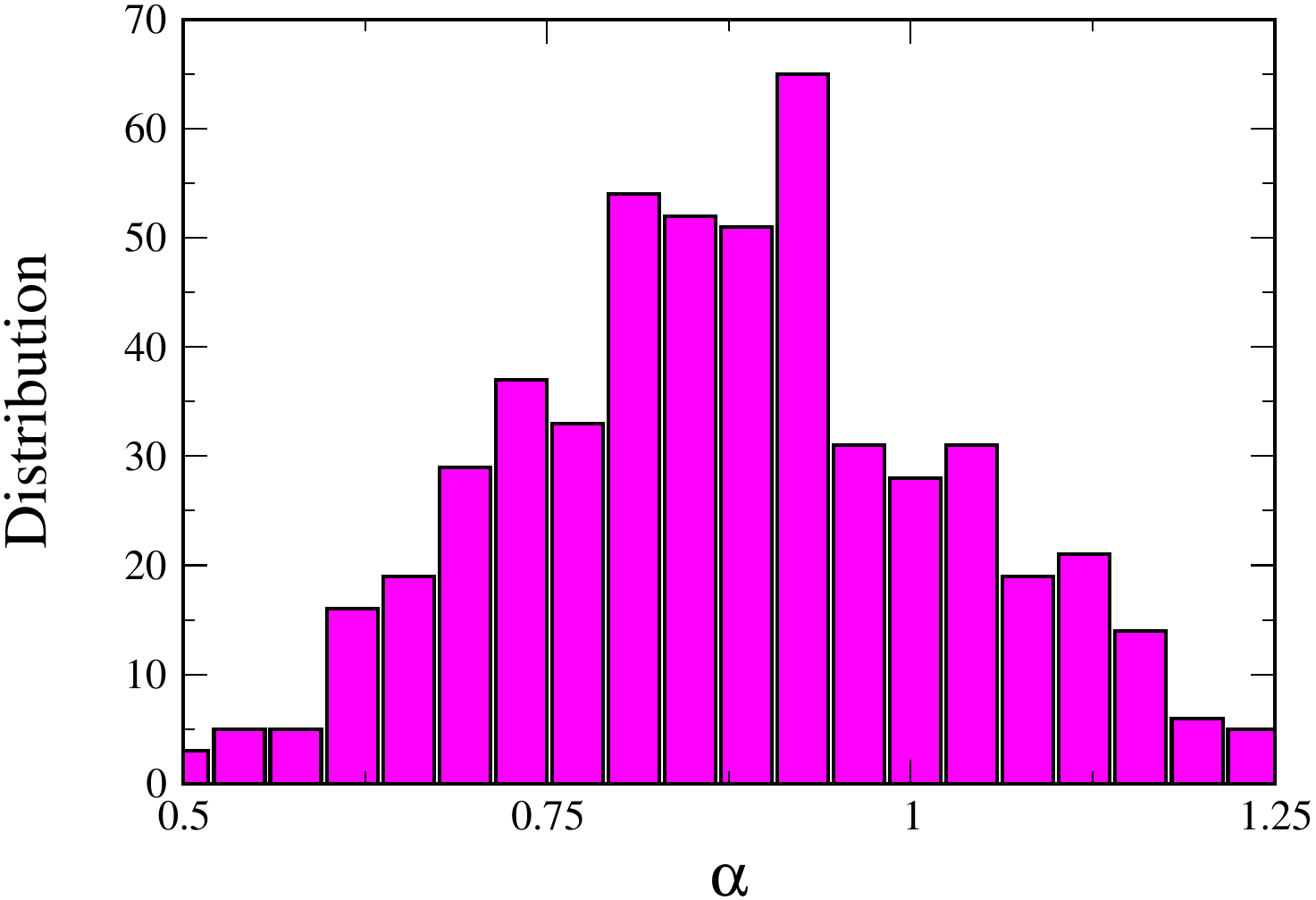}
(b)
\includegraphics*[width=7cm]{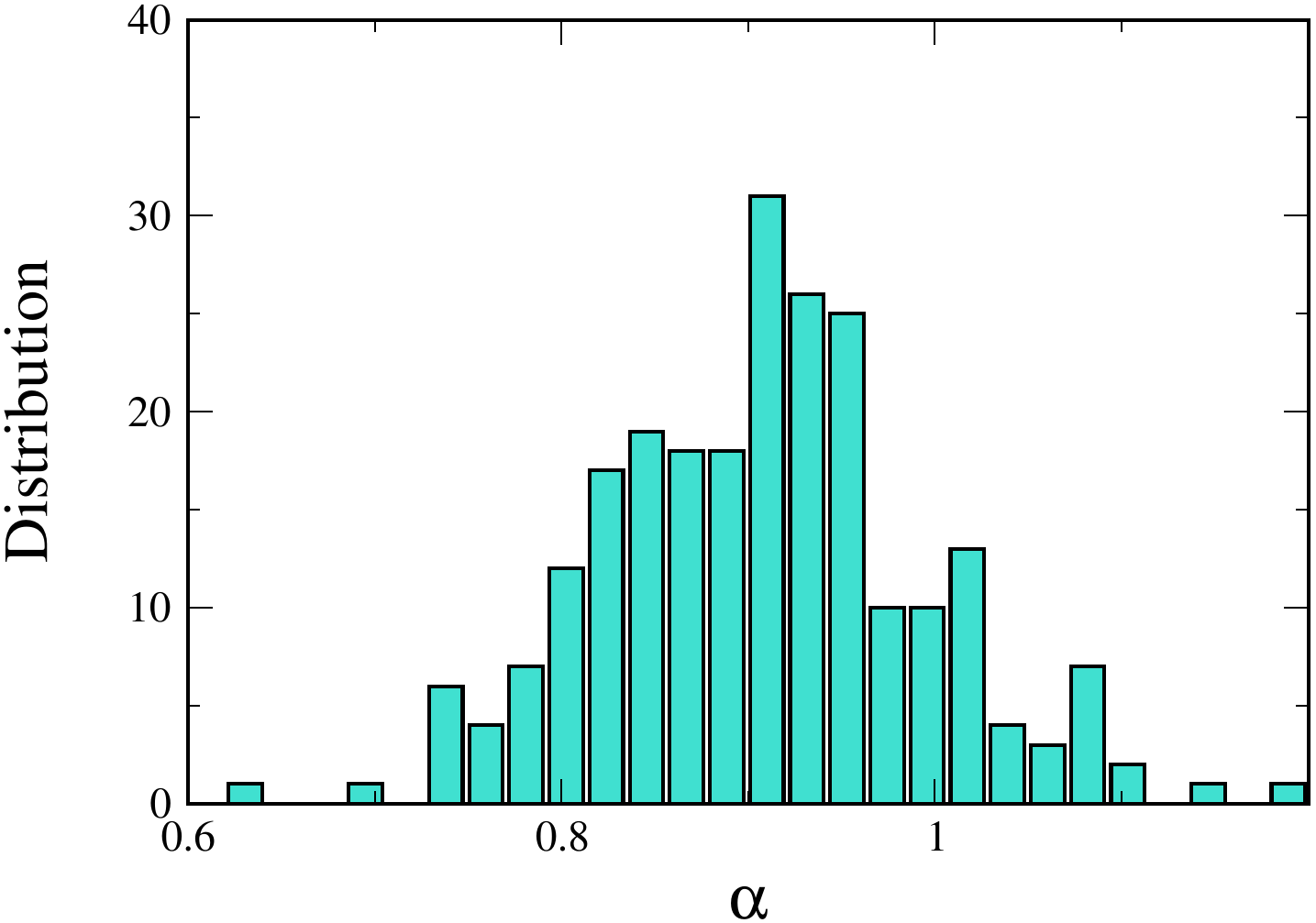}
}
\caption{(Color online) The distribution of the power of the power-law shaping of the microbranches arise nearby the main crack using (a) a perturbed honeycomb lattice  and (b) the  CRN model.}
\label{alpha1}
\centering{
(a)
\includegraphics*[width=7cm]{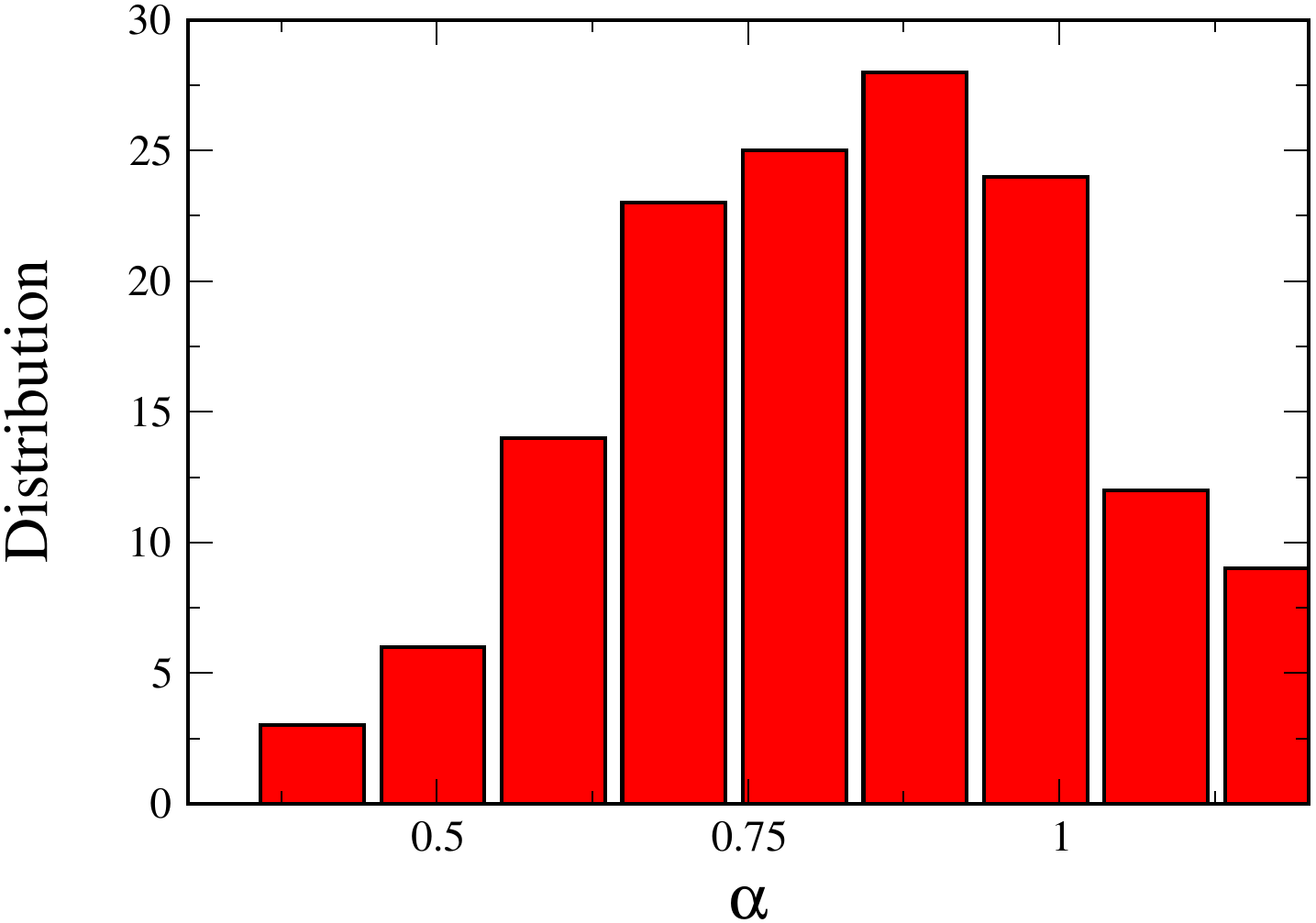}
(b)
\includegraphics*[width=7cm]{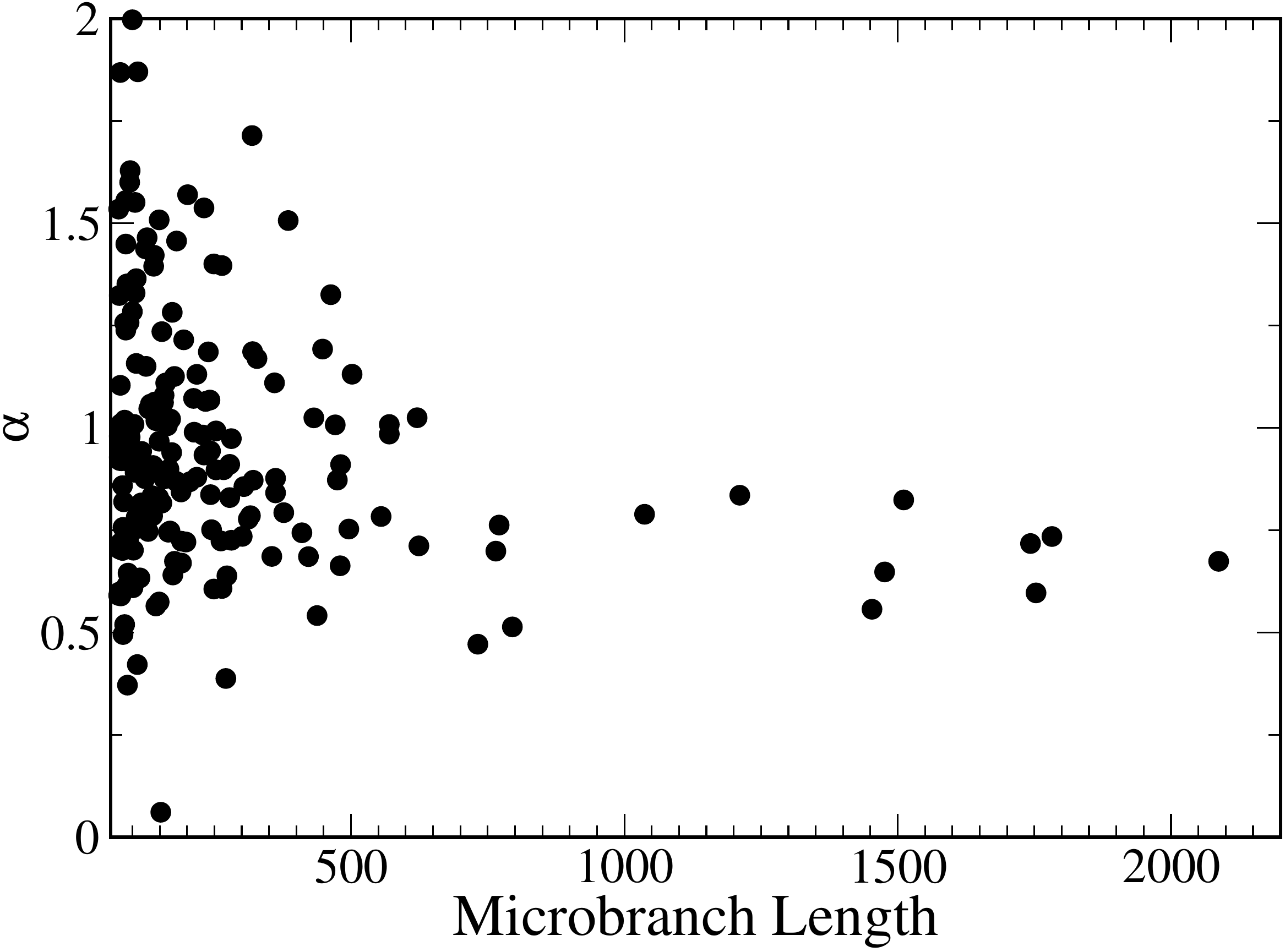}
}
\caption{(Color online) (a) The distribution of $\alpha$,  the power-law exponent  ($y=ax^{\alpha}$) of the microbranches arising near
the main crack for a perturbed hexagonal lattice. (b), A scatter-plot of the power $\alpha$ of the power-law exponent of the microbranches. We can clearly see that for the long branches,
$\alpha$ converges on the value $\alpha\approx0.7$, which is close to the experimental value. }
\label{alpha2}
\end{figure}

One of the most striking features that was discovered in the experiments was a power-law shape of the branches, $y=ax^{\alpha}$ ($x$ and $y$ are the spatial coordinates
of the microbranch such that $x=y=0$ is the location of the ``root" of the microbranch close to the main crack), with a universal power, $\alpha\approx0.7$
for several different materials~\cite{fineberg_sharon2,review}. 
On the one hand, our previous studies using CRN or a perturbed honeycomb lattice yielded only straight ($\alpha\approx1$)
microbranches~\cite{shay3,shay4}.
On the other hand, using a perturbed hexagonal lattice~\cite{shay4} we got branches that showed a power-law shape with
$\alpha\approx0.5$, though the results were noisy due to the relatively small number of broken bonds in each microbranch.
We note that the LEFM analysis of~\cite{katzav} yields $\alpha=2/3$, which is close to the experimental result,
but this analysis assumes symmetric branching, when our branching (and the experimental results) is by no means symmetric.
We also note that
the branches seen in the elastic beam model~\cite{beam1} or the Born-Maxwell model~\cite{bm2} look very noisy, without a clear power-law shape.
In this study, using larger lattices we can check the shape of the microbranches
based on relatively large microbranches ($\approx100$ broken bonds in each microbranch). In Figs. \ref{alpha1} and~\ref{alpha2}(a) we can see the power distribution for the different kinds of lattices.

\begin{figure}
\centering{\includegraphics[width=8cm]{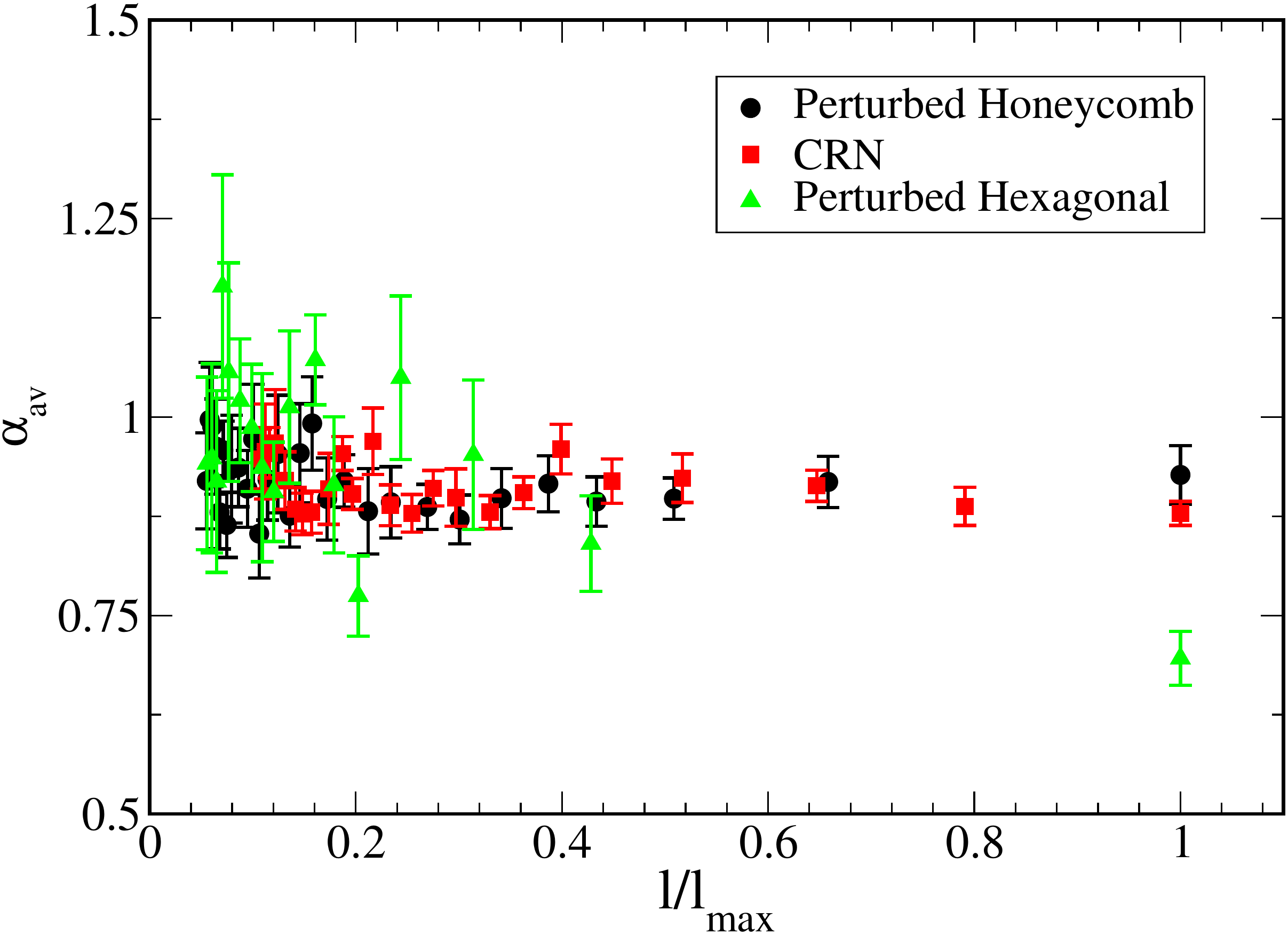}}
\caption{(Color online) The distribution of $\alpha$, the power-law exponent ($y=ax^\alpha$ of the microbranches arising near the main crack for the perturbed hexagonal, perturbed honeycomb and CRN models, as a function of the micro branch length.}
\label{alpha3}
\end{figure}

We can clearly see that the maxima of all the distributions are around $\alpha\approx0.85-0.9$, which is indeed less than 1 (straight lines). Looking closely at the larger lattices ($f=9$)
in Figs. \ref{pattarns} and \ref{pattarns2}  we can see the nonlinear power law shape of the large branches, especially
for the perturbed hexagonal lattice.
In Fig. \ref{alpha2}(b) we can see a scatter-plot of $\alpha$ as a function of the branch size for the perturbed hexagonal lattice case. It is clear that $\alpha$ converges on
the value $\alpha\approx0.7$ for the largest microbranches, which is close to the experimental value. The data for all three models is shown in Fig. \ref{alpha3}.  The other models do not show the decrease in $\alpha$ for the longest branches that is apparent for the hexagonal case,  and perhaps data for longer microbranches are needed in the other  cases, but in any case the asymptotic value of the exponent in all cases is less than unity.
\section{Discussion}
\label{discussion}

We have used relatively large lattices fracture simulations using GPU parallel computing (with $\order (5\cdot 10^6)$ particles)
for studying mode-I fracture. We find that the basic results from small
lattices ($\order (5\cdot 10^4)$ particles) are confirmed using the larger size systems. The fracture patterns look more physical with a larger system,
due to the large number of broken bonds in each microbranch. The width of the microbranch region relative to the system width decreases, a necessary condition if these models are to be taken seriously.  The basic  properties of the microbranches, like the
total length of the microbranches and the oscillations of the derivative of the electrical resistance with respect to the time based velocity measurements, which were extremely noisy when using small lattices, look more smooth and realistic in the larger
lattices. In particular there is now a clearer transition between the regime of steady state cracks, which there is a negligible amount of broken bonds beside the main crack, and the regime
of instability, where the amount of broken bonds in the microbranches increases dramatically. The sharp transition is particularly clear  in the hexagonal perturbed
lattice (Fig. \ref{hex_res}(b)).

In addition, important  features of the microbranches that have been found and studied experimentally, are recovered in our large system simulations. The 
correct branching angle is found, and in the CRN lattice  the correct variance is also obtained. The universal power-law shape that was found in different experimental studies~\cite{fineberg_sharon2}
is recovered here, and for the larger branches, we get the correct power in the hexagonal perturbed lattice.

In future work, we plan to exploit the power of GPU parallel computing to run 3D simulations using $\order (5\cdot 10^6)$ particles, with the goal of studying the different aspects concerning
the 3D nature of the microbranches. We intend to check the similarities and the differences between 2D and 3D Mode-I fracture simulations, and to find the regime when the 2D model is
sufficient, and on the other hand, the regime where 3D simulations are crucial.

\appendix

\section{GPU-accelerated $C$ Code}
\label{appendix_a}

In this appendix we discuss implementation of our codes using CUDA and the run-times of the CUDA runs using NVIDIA's Tesla C2050/C2070 GPU for the different 
modules that were used in this study,
and the acceleration ratios between a single CPU run and the GPU run. We re-wrote our $C$ codes and replaced the main time-demanding functions by CUDA kernels~\cite{padon}. 
CUDA is a computer code language that was developed by NVIDIA for using its CPU's. We note that in principle, one could use OpenCL,  a general language for any GPU device, though CUDA is optimized for 
NVIDIA's graphic cards.

For using GPU computing optimally, the code should be written as a function of the particular hardware. Naively, the computational tasks are divided into several blocks,  each block
containing a specific number of threads, when the threads in the same block execute simultaneously. In this work we used 512 threads per block which is the maximal number
of threads per block available on our graphic card. The parameters of the physical problem are loaded to the {\em global memory} of the GPU and each thread computes one computational
task, such as the force of one spring that acts on a pair of atoms. Such a naive choice however only produces up to a 10 times acceleration, since there are a lot of calls to the
global memory (which is relatively slow) for each atom calculating the force, especially in the 3-body force-law. Thus we extensively employed the option of using {\em shared memory}
(which is as fast as cache memory, shared for all the threads in the same block and limited to 64K) to further accelerate the simulations, especially in the molecular-dynamics module~\cite{padon,cooper}.
The global memory of Tesla C2050/C2070 GPU is about 2.5 Gbytes, which is the limiting factor of the number of atoms in  the 
 simulation. In this work we used double precision accuracy for all of our calculations (so using float precision will increase the number of atoms by a factor of 2, which is not
significant; the main issue of this study is the effect of increasing the number of atoms by two orders of magnitude).

The underlying plan of the main, molecular dynamics, module using shared memory is to split the (potentially random, and thus general) grid of atoms to several physical zones (very much like
Open-MPI as opposed to OpenMP parallelization) with lists that connect between the zones applied as a boundary condition for each physical zone; for the central elastic
force law (and also for the viscoelastic force law) we sort the bonds and for the 3-body force law we sort the atoms. Each zone (``block" in the CUDA-jargon) loads the locations and the velocities 
to a fast shared memory, and each ``thread" calculates the force of a certain bond (for the central force law), or atom (for the 3-body force law). Thus, instead of several calls to the global
memory for each atom's location, the calls are for the fast shared memory, and so good efficiency is then achieved.
Since the threads in the same block execute simultaneously, we have to use CUDA {\em Atomic} commands to sum correctly the contribution of the forces acts from the neighboring
atoms. Then, the new velocities and locations are calculated by simple CUDA kernels.

\begin{figure}
\centering{
\includegraphics*[width=7.5cm]{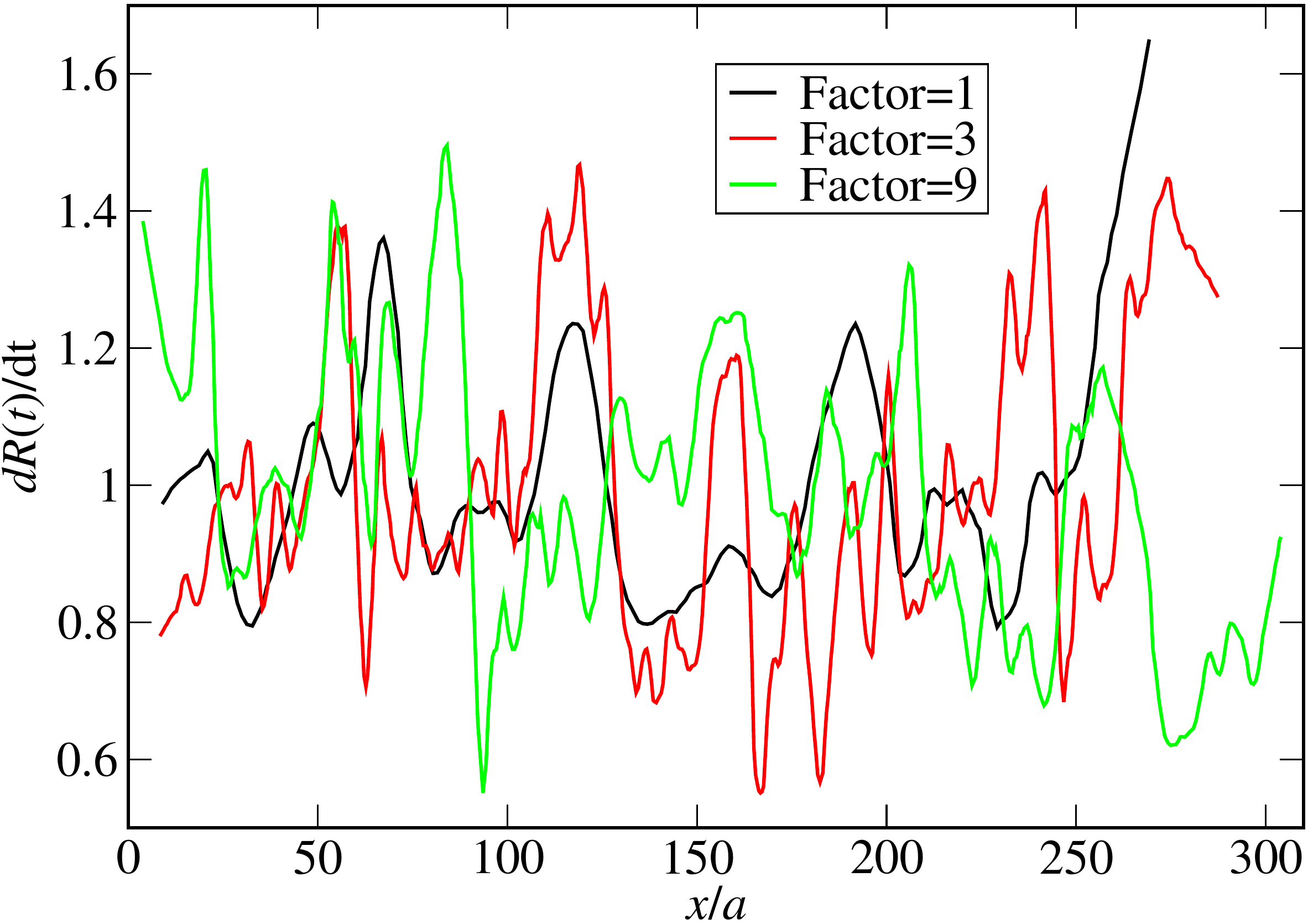}
}
\caption{(Color online) The derivative of the electrical resistance with respect to the time using a perturbed honeycomb lattice with $\nicefrac{\Delta}{\Delta_G}=3.4$ using
the different lattice sizes. The plots are normalized in the $x$-axis to the $f=1$ size (i.e. the $f=9$ results are divided by 9 etc.)}
\label{resis_demo}
\end{figure}

The electrical resistance module is basically solving a nonlinear Laplace equation. We used the methodology that was introduced in~\cite{bonamy} for calculating the electrical
resistance (a constant grid of bonds with $\sigma=1$), while a cracked bond in the molecular-dynamics module determines the ``cracked" ($\sigma=0$) bonds in the constant
grid of the electrical resistance. Thus, we used the same well-known methodology of using the CUDA kernels for solving 2D diffusion equations~\cite{cooper},
including the use of shared memory. We implemented
both Jacobi and red-black gauss-Seidel methods of solution, but no significant difference (in terms of the number of iterations for convergence) was found between the methods.
In Fig. \ref{resis_demo} we can see an example of the derivative of the electrical resistance with respect to the time using a perturbed honeycomb lattice for a specific
$\nicefrac{\Delta}{\Delta_G}$. We can see that using different size  lattices (using the new GPU code in the larger lattices and the CPU code in the smaller lattice),
we can see that the shape of the curves of electrical resistance look very much alike the experimental RMS
amplitude of the crack velocity~\cite{fineberg_sharon2,fineberg_sharon5} (that is measured via the electrical resistance~\cite{shay3}).

The parallel Monte-Carlo algorithm module for generating the CRN required extra care. Since each possible switch of bonds should be considered energetically independently, each
switch should be separate from all the other simultaneous possible switches (to avoid over-lapping of the switches and their neighboring zone).  We mention that we use the parallel {\em THRUST} library (in CUDA) for sorting efficiently the
nearest bonds for each bond, every  given number of cycles.

\begin{figure}
\centering{
(a)
\includegraphics*[width=7cm]{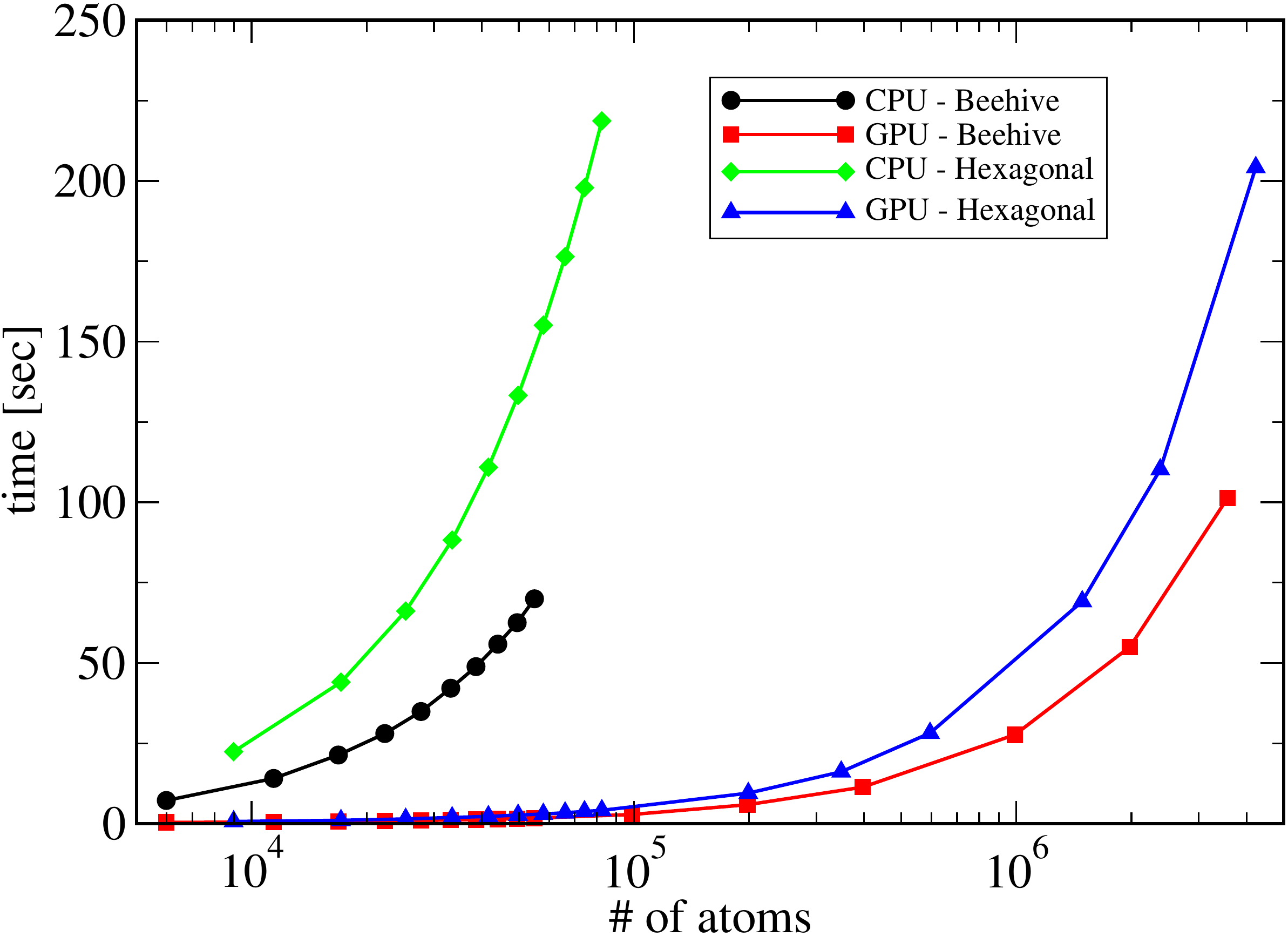}
(b)
\includegraphics*[width=7cm]{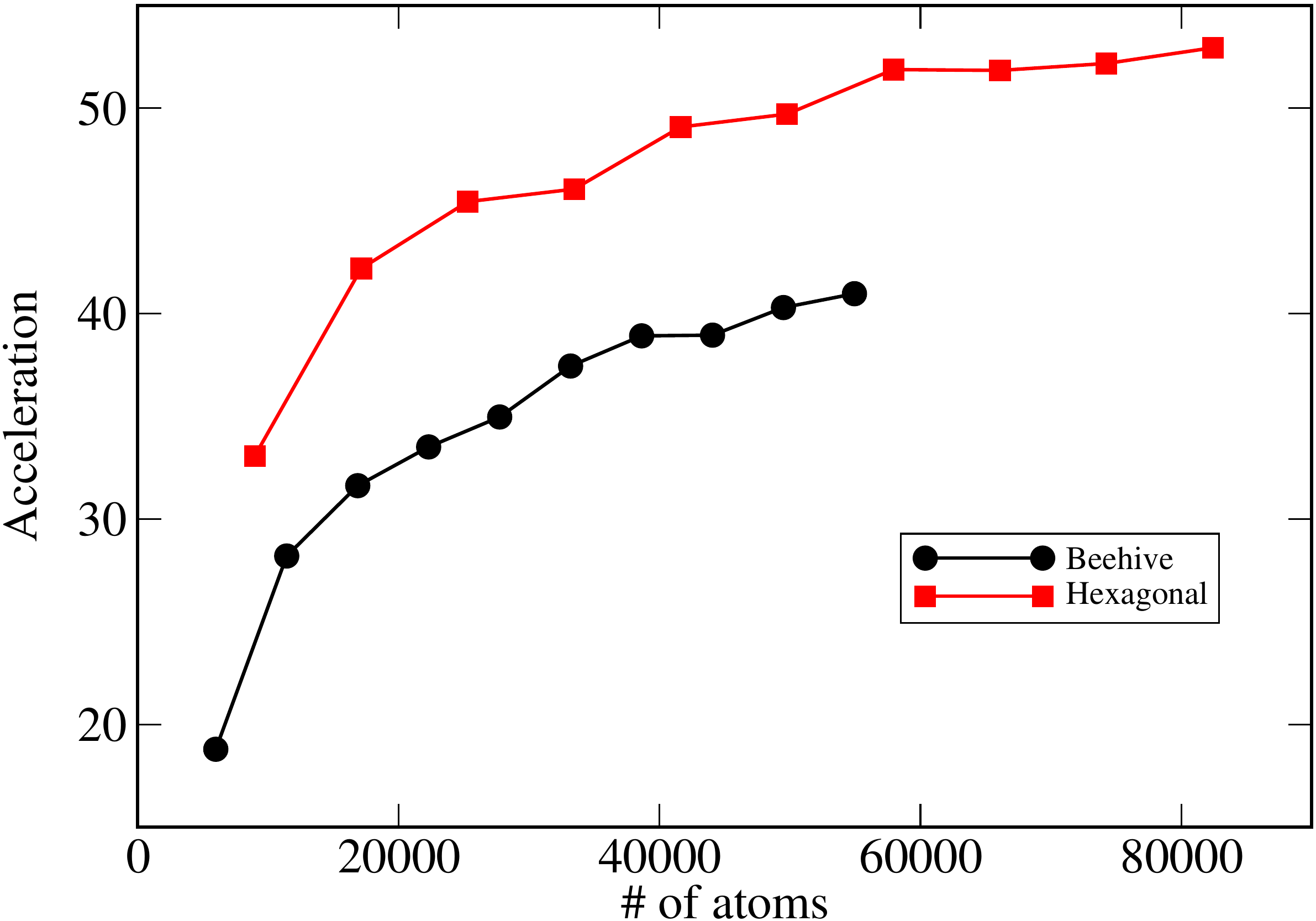}
}
\caption{(Color online) (a) Simulation times (in $[\sec]$) of 1000 cycles of the molecular-dynamics Euler scheme as a function of the system size (both using honeycomb and hexagonal lattices)
using unoptimized $C$ code with CPU and with CUDA using GPU.
(b) The acceleration run times between GPU and CPU as a function of the system size  for honeycomb and hexagonal lattices).}
\label{times_md}
\end{figure}
\begin{figure}
\centering{
(a)
\includegraphics*[width=7cm]{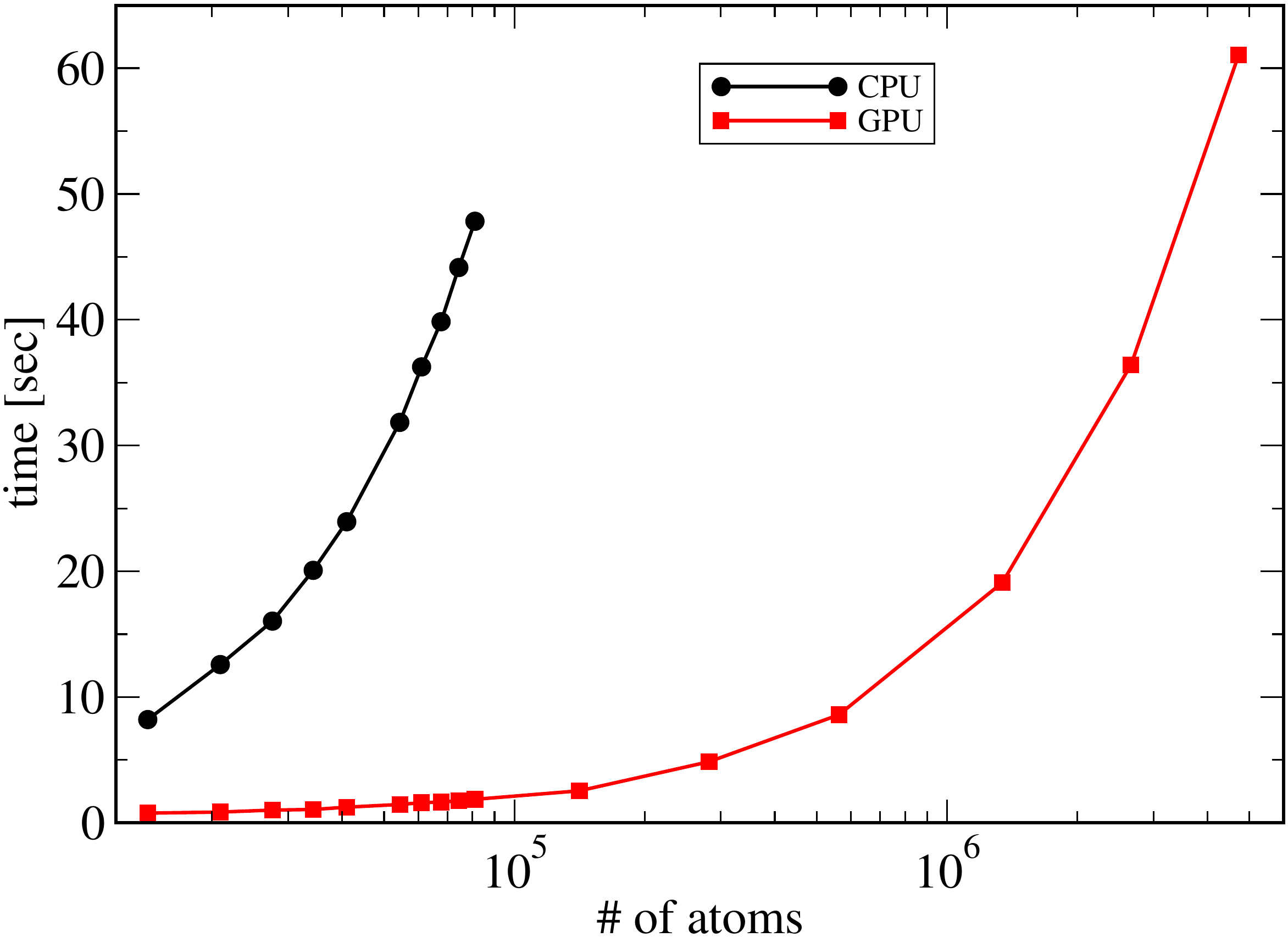}
(b)
\includegraphics*[width=7cm]{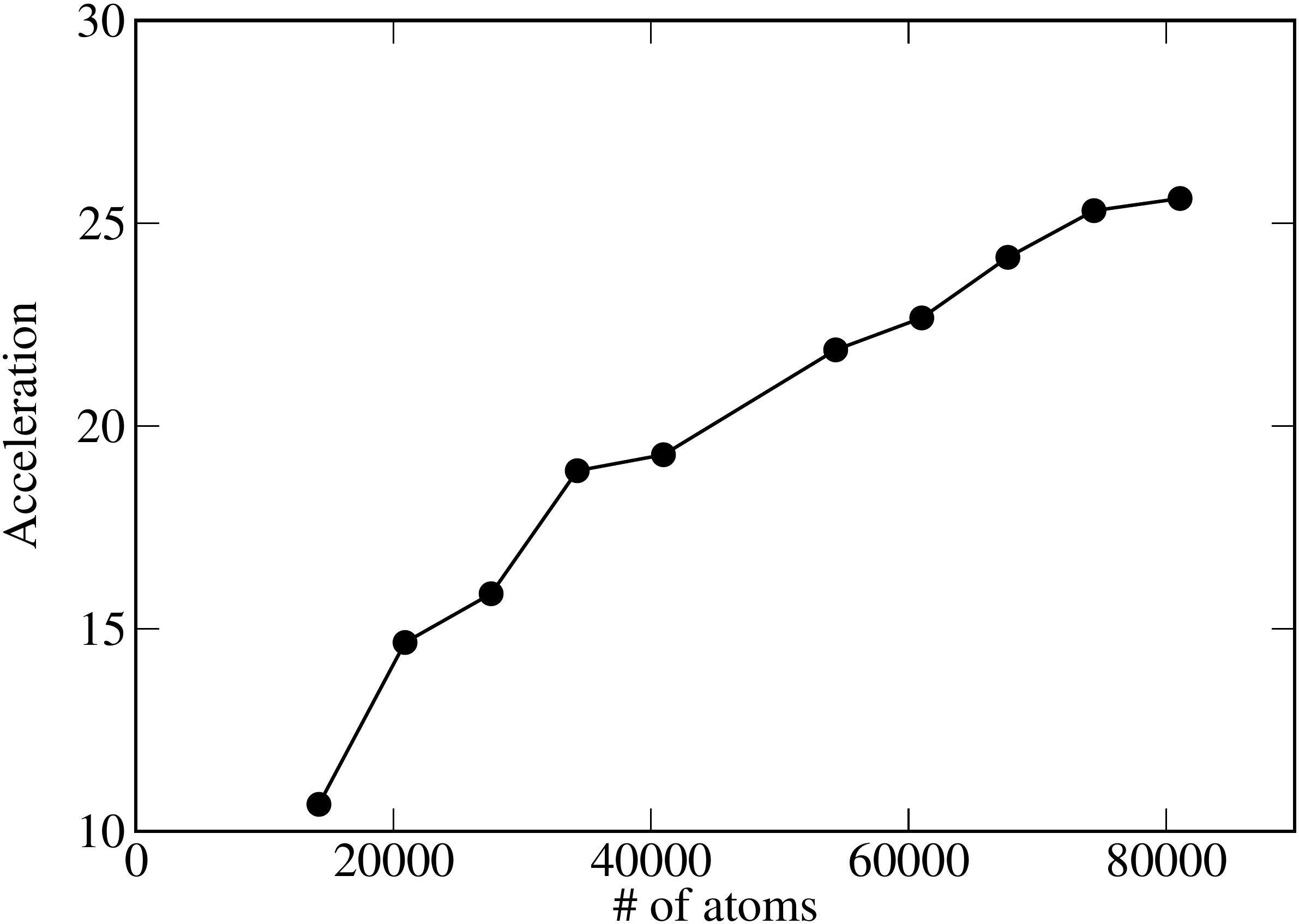}
}
\caption{(Color online) (a) Simulation times (in $[\sec]$) for 1000 iterations of a Jacobi-method Laplace solver (for the electrical resistance) as a function of the system size
using unoptimized $C$ code with CPU and with CUDA using GPU (the times are similar also in Red-Black Gauss Seidel method).
(b) The acceleration run times between GPU and CPU as a function of the system size.}
\label{times_lap}
\end{figure}

In Figs. \ref{times_md}-\ref{times_lap}(a) we can see the typical run times for 1000 cycles (in seconds) for the molecular dynamics module and the Laplace solver module as a function
of the system size. We can clearly see the benefit of using GPU computing, while the main benefit is the possibility to run systems with large number of atoms, which with a single
CPU, was prohibitively time-consuming. The run-time using $\order(10^4)$ atoms with a single CPU is similar to the run-time using $\order(10^6)$ atoms with a single GPU.
In Figs. \ref{times_md}-\ref{times_lap}(b) we see the acceleration ratios between a single CPU and a single GPU (of course, only for small systems, when a CPU
run is available). We can see the significant acceleration, 40 times faster for the honeycomb lattice and over 50 times faster
for a hexagonal lattice due to more demanding 3-body force law). In the Laplace solver the speedup is a little bit lower and stands at approximately 25 times faster in GPU versus a single CPU.

\begin{figure}
\centering{
(a)
\includegraphics*[width=7cm]{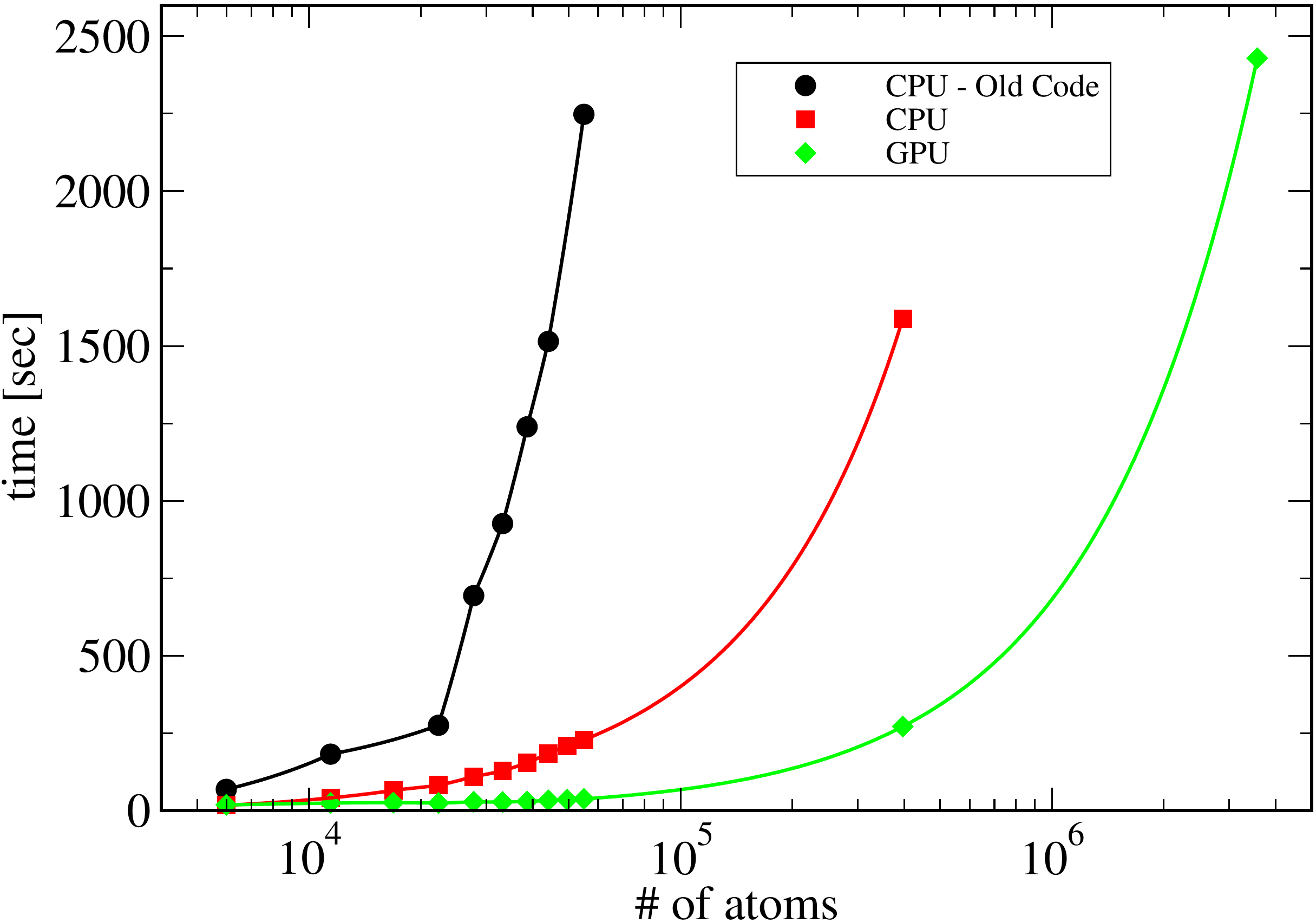}
(b)
\includegraphics*[width=7cm]{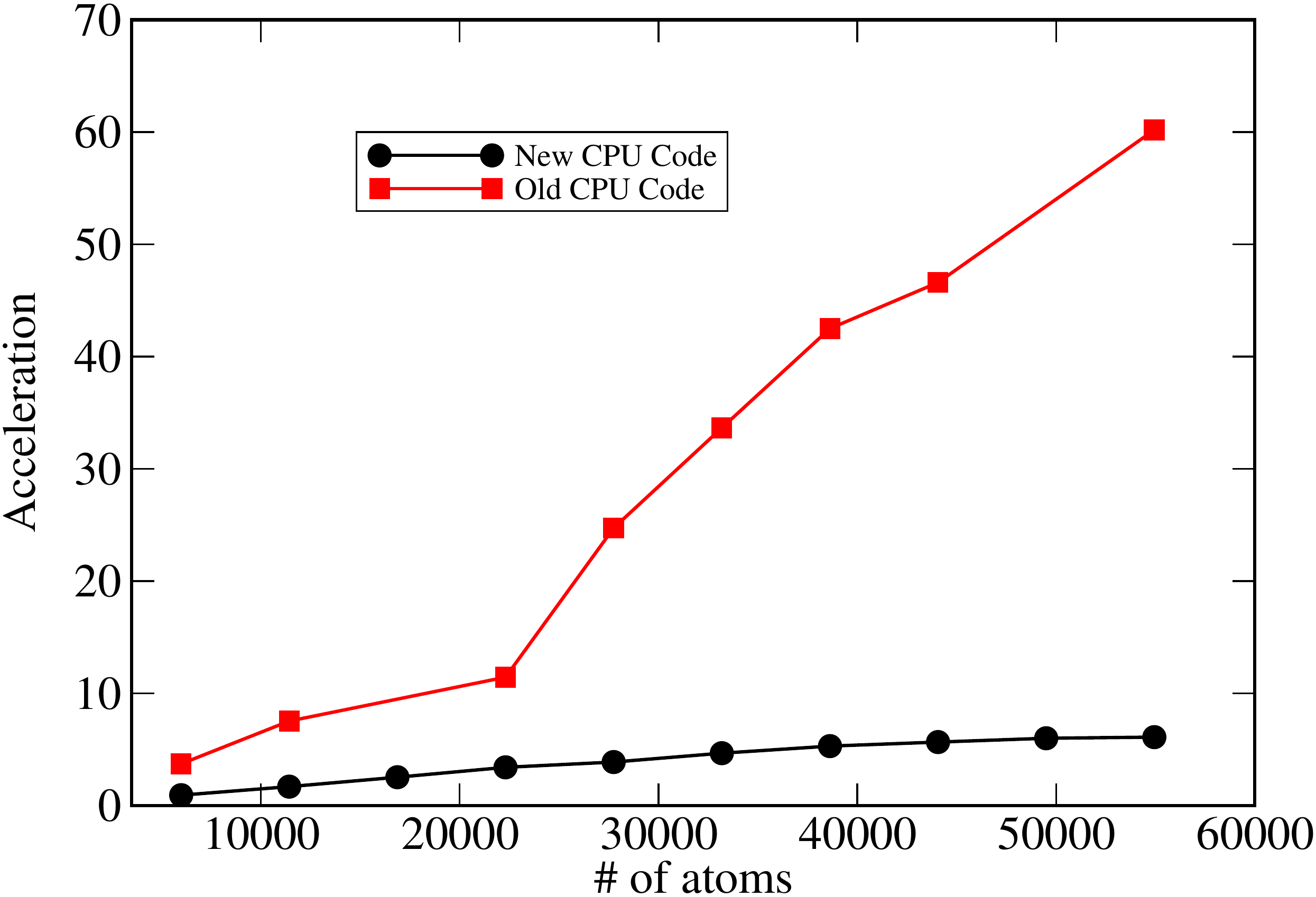}
}
\caption{(Color online) (a) Simulation times (in $[\sec]$) for a WWW Monte-Carlo scheme for producing a CRN (simulation stopped while the energy reaches half of the initial random energy)
as a function of the system size
using unoptimized $C$ code with CPU, with CUDA using GPU and with the ``old-CPU" code used in~\cite{shay3}.
(b) The acceleration run times between GPU and CPU and the ``old-CPU" codes as a function of the system size.}
\label{times_mc}
\end{figure}
In Fig. \ref{times_mc} we can see the run-times using a single CPU and the GPU. We can see here that the acceleration ratio here is lower then in the previous modules (about
$\approx5-10$), but still, in larger lattices (of $\order(10^6)$ atoms), the benefit is clear. We mention that since the programing using CUDA is much more demanding from
the programmer, especially regarding the memory management, while re-programing the code, we improved our old-CPU code (that was in use in~\cite{shay3}),
significantly; the acceleration ratio of the GPU code to the old CPU code is $\approx 60$.

\section{CRN Monte-Carlo parallel CUDA algorithm}
\label{appendix_b}

\begin{figure}
\centering{
(a)
\includegraphics*[width=7cm]{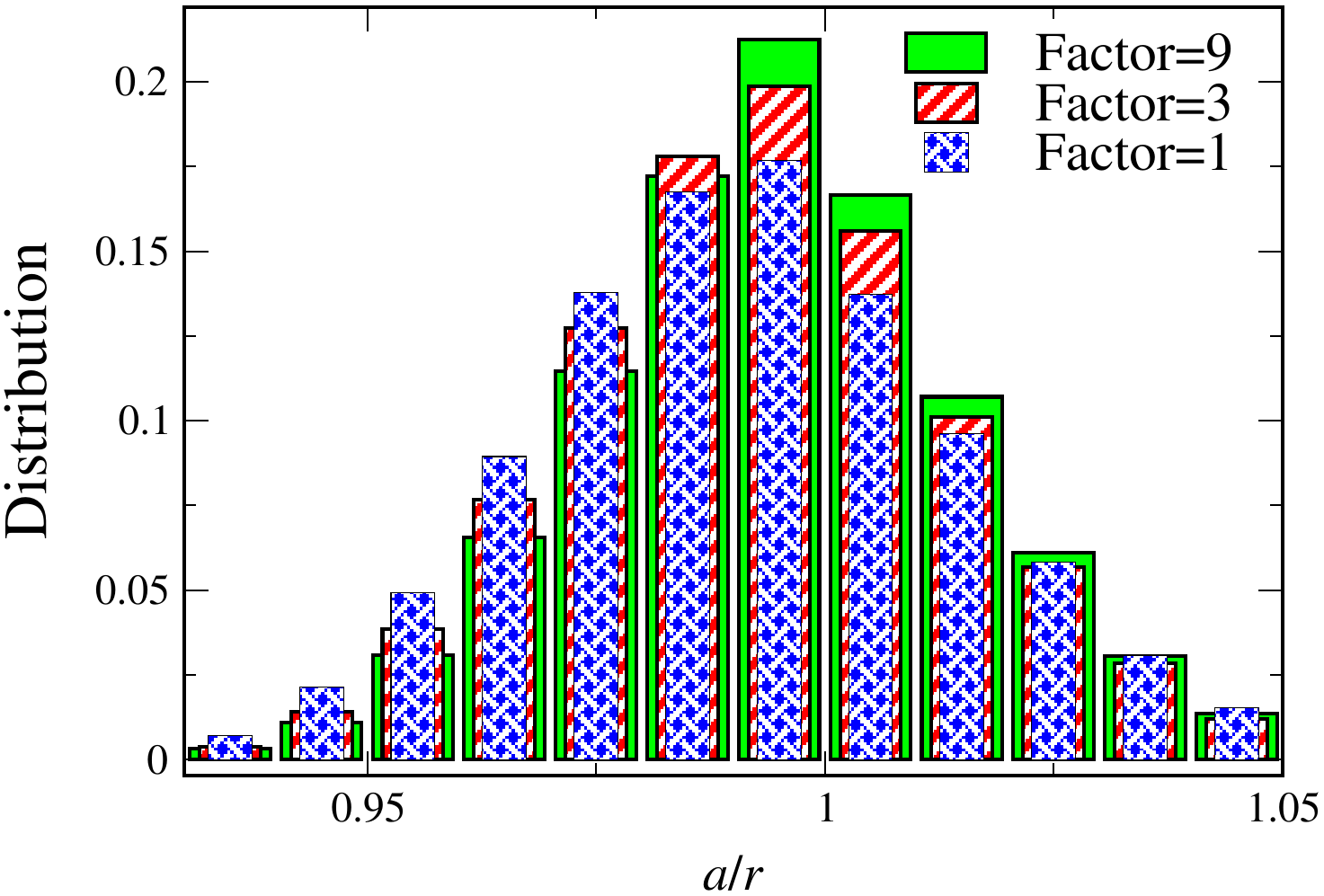}
(b)
\includegraphics*[width=7cm]{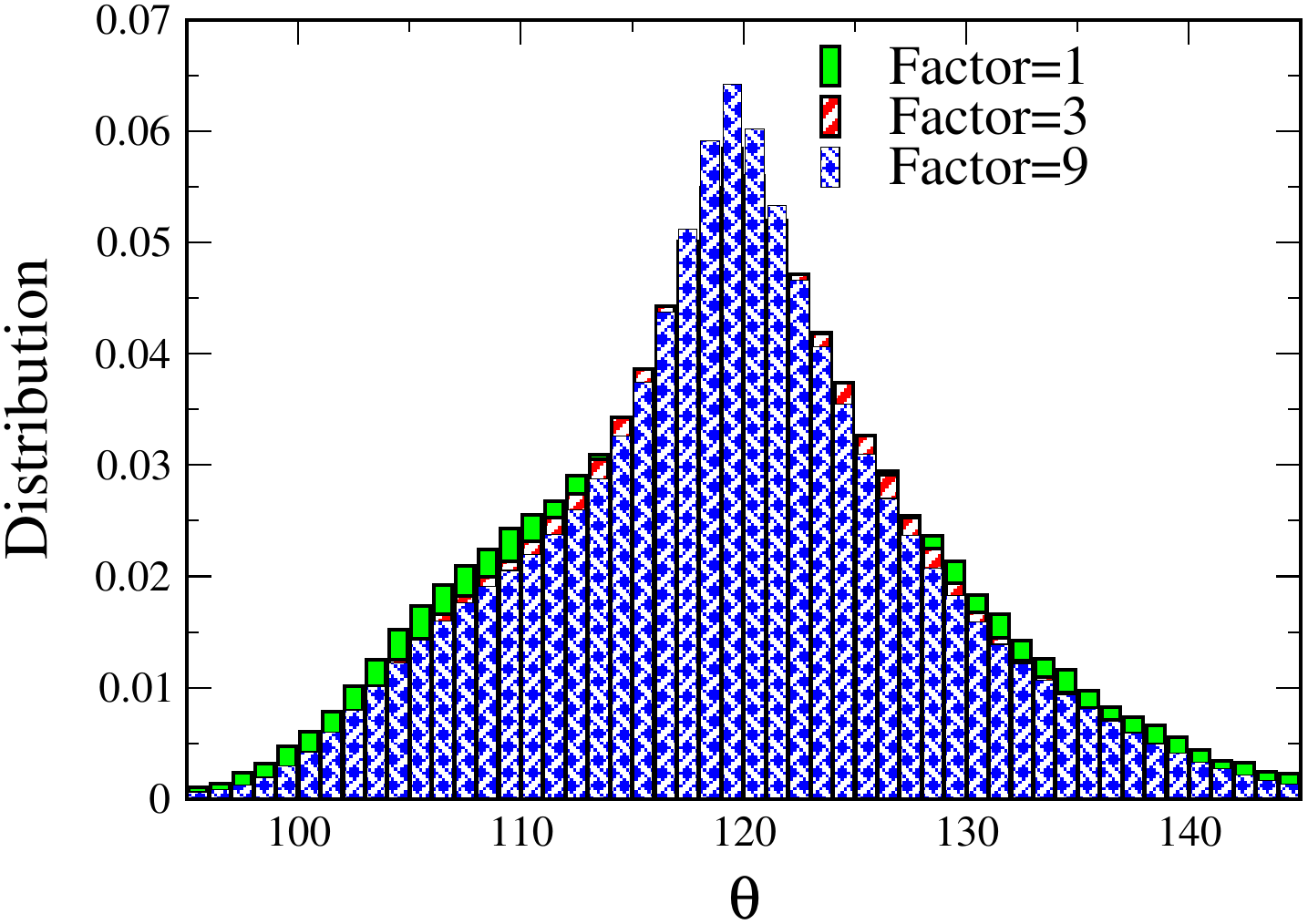}
}
\caption{(Color online) (a) The radial distance distribution of the bonds in CRN using different lattice size. (b) The angular distribution of the bonds in the CRN using different lattice size.}
\label{crn_histograms}
\end{figure}

\begin{figure}
\centering{
\includegraphics*[width=7cm]{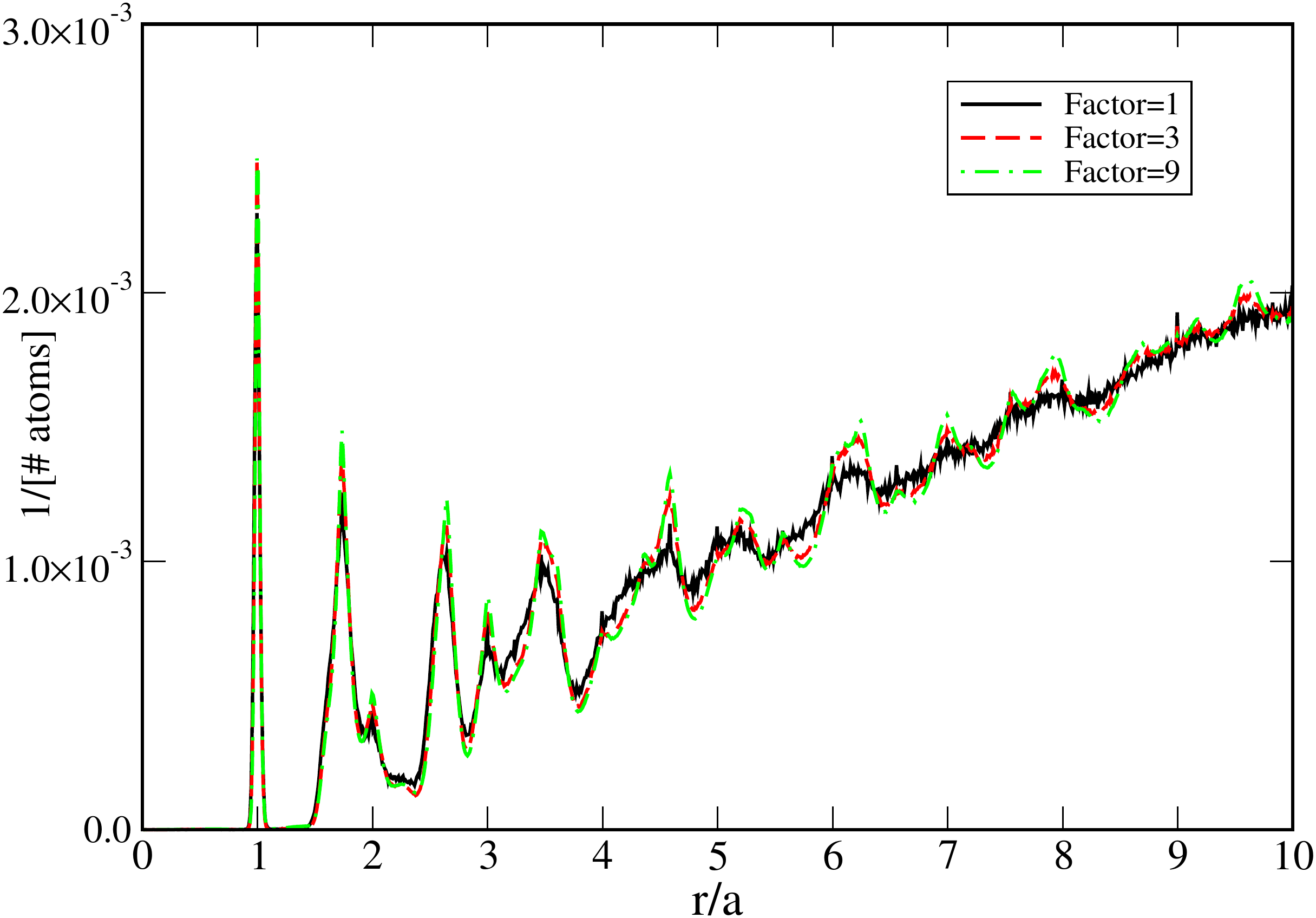}
}
\caption{(Color online) The radial distribution function (RDF or $g(r)$) of the the CRN using different size of lattices.}
\label{rdf}
\end{figure}

In previous work~\cite{shay3} we have shown that the CRN shares similar features with  real amorphous matter, like amorphous silicon~\cite{laaziri}. In this appendix
we check explicitly the quality of the parallel GPU algorithm, generating a CRN. In Fig. \ref{crn_histograms} we can see the radial and angular distributions of the bonds in the CRN
using different lattice size, and in Fig. \ref{rdf} we can see the radial distribution function $g(r)$ using different lattice size.
$f=1$ (of $\order(10^4)$ atoms) is the data using the old CPU code that was in use in~\cite{shay3}, when larger lattices was produced via the new GPU algorithm. 

We can see that the radial and angular distributions and the $g(r)$ curves are similar under the scaling of the lattice size (the number of the bonds or angles). This proves
the validity of the parallel GPU algorithm, comparing the old CPU algorithm (that was verified before against experiments).
Moreover, the RDF in the larger lattice size is smoother due to better statistics.  

\section{The Rayleigh surface wave speed with $k_{\theta}\ne 0$ lattices}
\label{rayleigh}

Since the models in this paper use a 3-body potential law (aside by the central two-body force law) with $k_{\theta}\ne0$, we need to recalculate the 
Rayleigh wave speed $c_R$, which is the terminal velocity for mode-I fracture~\cite{freund} for different $k_{\theta}/k_r$. The most convenient way to calculate
the Rayleigh wave speed is to calculate first the longitude (primary) $c_l$ and the transverse (secondary) $c_t$ wave speeds and then, to calculate the
Rayleigh wave speed via the well-known formula~\cite{achenbach}:
\begin{equation}
\left(1-\frac{c_R^2}{c_t^2}\right)^2-4\left(1-\frac{c_R^2}{c_l^2}\right)^{\nicefrac{1}{2}}\left(1-\frac{c_R^2}{c_t^2}\right)^{\nicefrac{1}{2}}=0
\label{cr}
\end{equation}
For $k_{\theta}=0$ the expressions for $c_l$ and $c_t$, and thus also for $c_R$, can be derived analytically for a 2D hexagonal lattice with $a$ as the lattice scale
(for $k_r=m=1$; the wave velocities scales as $\sqrt{k_r/m}$):
\begin{equation}
c_l^{\mathrm{Hex}}=\frac{3}{\sqrt{8}}a,\qquad c_t^{\mathrm{Hex}}=\sqrt{\frac{3}{8}}a,\qquad c_R^{\mathrm{Hex}}=\frac{\sqrt{3-\sqrt{3}}}{2}a,
\label{wave_hex}
\end{equation}
and for a 2D honeycomb lattice:
\begin{equation}
c_l^{\mathrm{Hon}}=\frac{3}{4}a,\qquad c_t^{\mathrm{Hon}}=\frac{\sqrt{3}}{4}a,\qquad c_R^{\mathrm{Hon}}=\frac{\sqrt{3-\sqrt{3}}}{2\sqrt{2}}a
\label{wave_hon}
\end{equation}

\begin{figure}
\centering{
(a)
\includegraphics*[width=7cm]{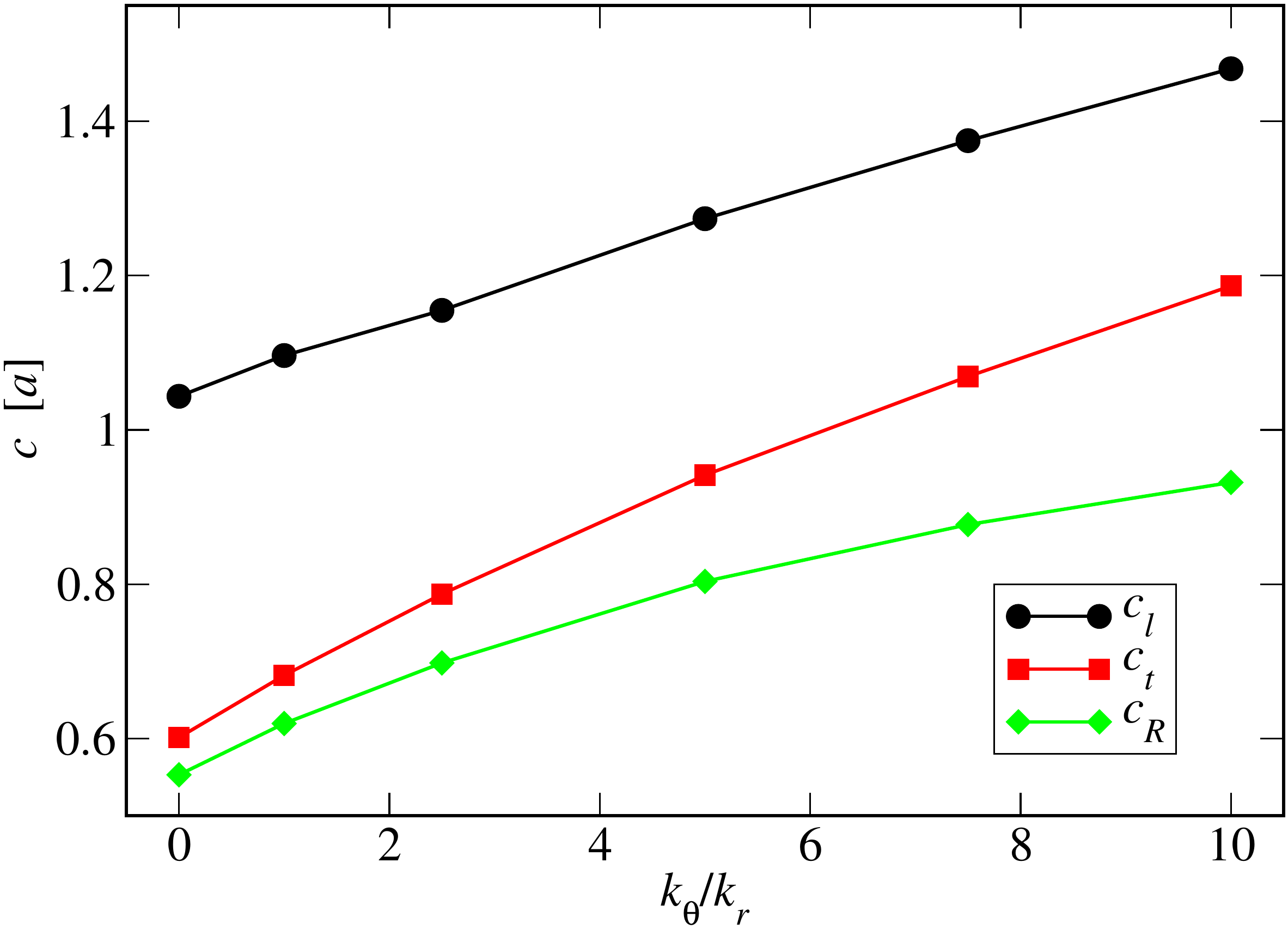}
(b)
\includegraphics*[width=7cm]{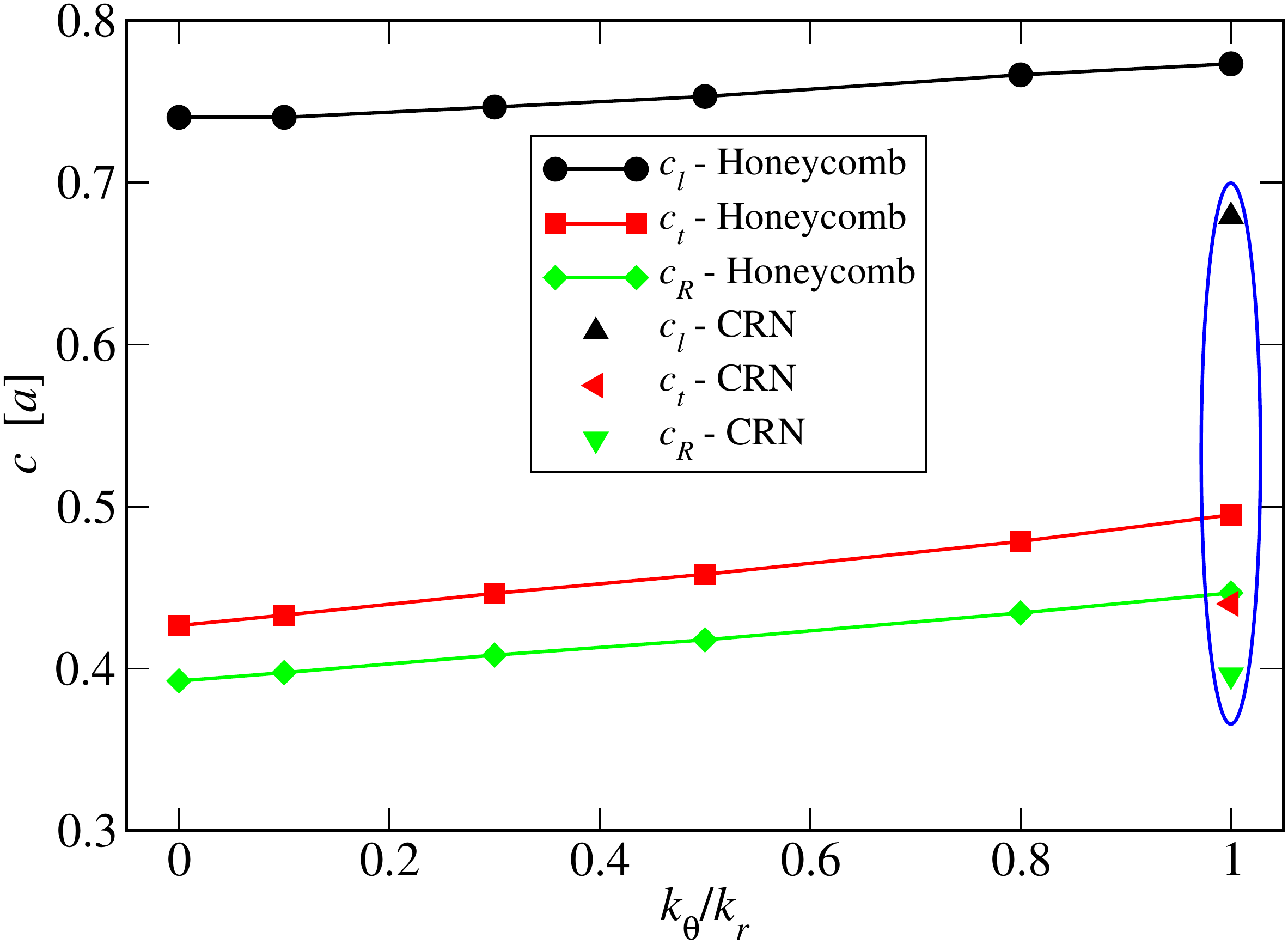}
}
\caption{(Color online) (a) The longitude and the transverse sound waves speeds along with the resulting calculated Rayleigh surface wave speed using Eq. \ref{cr} for the hexagonal lattice as a function
of $k_{\theta}/k_r$. (b) The same in the honeycomb lattice. The sounds velocities for the CRN that was used in this paper ($k_{\theta}/k_r=1$) are presented In the ellipse.}
\label{sounds}
\end{figure}

We calculate $c_l$ and $c_t$ via measuring the wave velocities by initiating longitude and transverse small deformations in the end of the samples in the different
lattices that we use in this study and then find $c_R$ via Eq. \ref{cr}. The results are shown in Fig. \ref{sounds}(a) for the hexagonal lattice and in Fig. \ref{sounds}(b)
for the honeycomb and CRN lattices.

We can see that for both lattices, the numerical value for the wave velocities using $k_{\theta}=0$  reproduce the analytical values, Eqs. \ref{wave_hex} and \ref{wave_hon}, respectively.
In the hexagonal lattice using $k_{\theta}/k_r=10$ (which was the value used in this study) the Rayleigh wave speed increases by $\approx65\%$ relative to the $k_{\theta}=0$ value. In the honeycomb
lattice, where we use $k_{\theta}/k_r=1$ at most, the Rayleigh wave speed increases by $\approx12\%$ relative to the $k_{\theta}=0$ value. We note the CRN speeds (in the ellipse
in Fig. \ref{sounds}(b)) are a little bit slower than the  pure honeycomb lattice. In addition we note that the random noise of the perturbed lattice changes the wave speeds less than $1\%$, and that
the wave speeds are not affected at all by $\alpha_{\mathrm{pot}}$.

\begin{acknowledgments}
S.I. Heizler and D.A. Kessler wish to thank Jay Fineberg and his research group in HUJI for useful advice and comments.
S.I. Heizler wishes to thank Oded Padon from TAU for helpful advises and lessons in using CUDA and GPU computing.
\end{acknowledgments}

\end{document}